\documentclass[english,12pt]{article}
\usepackage[T1]{fontenc}
\usepackage[utf8]{inputenc}
\usepackage[english]{babel}
\usepackage[pdftex]{graphicx}
\usepackage[width=\textwidth]{caption}
\usepackage[top=21mm, bottom=21mm, left=24mm, right=24mm]{geometry}
\usepackage{array,float, placeins,flafter, longtable,booktabs,pdflscape}
\usepackage{verbatim}
\usepackage{lmodern}
\usepackage{xr}
\usepackage{mathtools}
\usepackage{multirow}
\usepackage{changepage}
\usepackage{xurl}   
\usepackage{threeparttable}
\usepackage{dcolumn}
\newcolumntype{L}{D{.}{.}{2,3}}
\usepackage{amsfonts,amsmath,amsthm,amssymb,bm,bbm,dsfont}
\usepackage{lmodern}
\usepackage{enumitem}
\usepackage{natbib,color}
\usepackage{subfigure,float,booktabs}
\usepackage{multicol}
\definecolor{winered}{rgb}{0.5,0,0}
\usepackage[pdfstartview=FitH, bookmarks=true, bookmarksopen=true,
breaklinks=true, colorlinks=true, linkcolor=blue,
anchorcolor=my, citecolor=blue, filecolor=blue,
menucolor=blue, urlcolor=black]{hyperref}

\numberwithin{equation}{section}
\theoremstyle{definition}

{\theoremstyle{plain}
	}
\definecolor{my}{rgb}{0.05,0.05,0.5}
\definecolor{myBlue}{rgb}{.1,.1,.5}
\definecolor{myGreen}{rgb}{0,.4,0}
\definecolor{myRed}{rgb}{.25,0.15,.5}

\newcommand{\cond}{\displaystyle \stackrel{d}{\longrightarrow}}

\renewcommand{\mathbf}[1]{\textbf{\textit{#1}}}

\newcommand{\Rmnum}[1]{\expandafter\@slowromancap\romannumeral #1@}
\allowdisplaybreaks
\usepackage{abstract}

\makeatletter
\g@addto@macro\maketitle{\vspace{-2em}}
\makeatother

\begin{document}
	\title{Digital Divide: Evidence from the 2020
		Canadian Internet Use Survey%
		\thanks{\scriptsize This version: \today. The authors thank Henry Kim, the participants
			of the 13th CEQURA, 48th ACEA, 57th CEA, and 38th CESG Conferences, and the
			seminar of CBDC Econ Working Group at the Bank of Canada for
			helpful comments. We acknowledge financial support of the
			Catalyzing Interdisciplinary Research Clusters Initiative, York
			University. This research is part of the project ``Digital
			Currencies'', approved and funded by the office of VPRI and
			participating faculties of York University. The authors acknowledge
			access to the data provided by Statistics Canada Research Data
			Centres (Project 21-MAPA YRK-721). The second author was awarded
			Best Student Presentation at the 48th ACEA Conference.}}
	
	\date{}
	
\author{
	Joann Jasiak\thanks{\scriptsize York University,
		\texttt{jasiakj@yorku.ca}}
	\and
	Peter MacKenzie\thanks{\scriptsize York University,
		\texttt{petem9@yorku.ca}}
	\and
	Purevdorj Tuvaandorj\thanks{\scriptsize York University,
		\texttt{tpujee@yorku.ca}}
}

	\maketitle
	
\begin{abstract}
	This paper studies inequality in digital participation across
	socioeconomic and demographic groups using the 2020 Canadian
	Internet Use Survey (CIUS). We combine survey-weighted logistic Lasso,
	an exact Shapley decomposition of age--education gaps, a
	sequential logit, and a bifactor item response theory (IRT) measure of digital literacy
	to identify who is excluded, why gaps persist, and where along
	the adoption path they arise. 
	
	Education is the only determinant
	that remains significant at every rung of the digital ladder.
	Income inequality is most pronounced for virtual-wallet adoption;
	for online banking, employment and education together account for
	nearly half of the pro-rich concentration, indicating a broad
	socioeconomic gradient rather than a purely income-based divide.
	Persons with disabilities face the largest penalty at the
	digital-payments stage rather than at online banking, pointing
	to accessibility gaps in retail payment interfaces. Conditioning
	on digital literacy eliminates the education gradient at internet
	entry and reduces it by 61\% at the online banking rung, but a
	substantial residual persists, pointing to behavioral and
	institutional frictions beyond measurable competence. The
	youngest cohort records the lowest information-seeking score
	despite high digital engagement, and security deficits are
	concentrated among landed immigrants and visible minorities.
	\begin{description}
		\item[Keywords:] Digital inequality, income concentration,
		logistic Lasso, survey microdata, sequential logit,
		item response theory.
		\item[JEL codes:] C55, D63, I24, O33.
	\end{description}
\end{abstract}
	
\section{Introduction}
\label{sec:intro}

\textit{What we know about the Canadian digital divide.}
There is broad agreement that digital participation in Canada is
unequally distributed along familiar socioeconomic lines.
\citet{Haight-et-al(2014)}, using the 2010 CIUS, show that age,
income, and education are the primary predictors of internet use and
that the digital divide extends well beyond infrastructure access to
encompass the skills and willingness needed for meaningful engagement.
\citet{Wavrock_2022} document significant differences in the breadth
and depth of digital engagement across age, education, and income
groups in the 2018 CIUS, while administrative sources show that the
urban--rural gap in broadband access remains substantial despite
federal investments under the Universal Broadband Fund
\citep{CRTC2023, ISED2024}. The resulting picture is one of layered
disadvantage: low-income households, older adults, and individuals
with limited educational attainment are less likely to access the
internet and, conditional on access, less likely to use it for
financially or socially consequential activities
\citep{VDVD2019, TheDais2024}.

\noindent\textit{What distinguishes Canada.}
Canada's digital inequality challenge has features that set it apart
from other developed economies. The central issue is not absent
infrastructure: by 2020, 92.2\% of CIUS respondents reported
using the internet. Instead, inequality reflects differences in the
cost, reliability, and quality of connections across geographic areas;
the absence of a coordinated national digital literacy strategy
(Canada remains one of the few OECD countries without one); and the
concentration of exclusion among specific populations --- Indigenous
communities, persons with disabilities, low-income seniors, and recent
immigrants \citep{TheDais2024, ESDC2024, Kwok-Korpela-2024} --- who
are also among those most reliant on digital services to access health
information, government benefits, and financial services as physical
alternatives continue to decline.
	
\paragraph*{Four gaps this paper addresses.}
Despite a growing descriptive literature, four questions remain
unanswered or only partially answered.

\begin{enumerate}[leftmargin=1.5em, label=\textbf{Q.\arabic*.}]
	\item \textbf{Which characteristics predict digital financial service
		adoption once the full covariate space is taken into account?}
	Most existing studies rely on a small, fixed set of regressors. In
	CIUS 2020, the set of plausible predictors is broad, including age,
	education, income, employment, language, household composition,
	immigration status, Indigenous identity, visible minority status,
	disability, health, and province. Because these variables are categorical, dummy expansion
	yields a relatively large candidate regressor set, making
	standard stepwise selection ad hoc and uncorrected
	post-selection inference invalid.
	
\item \textbf{How much of the age--education gap in digital financial
service use is explained by observable mediators?}
Most studies identify \emph{who} is excluded but say little about
\emph{why}. Differences in online banking or digital payment use may
reflect income constraints, health and disability limitations, or
personal preferences rather than digital skill or access barriers per se. Distinguishing among these channels is essential for targeting policy effectively.
	
	\item \textbf{At which stage of the adoption process do different
		groups fall behind?}
	The CIUS routes respondents sequentially: internet users are asked
	about online banking, and those who shop online are asked about
	payment methods. A lower unconditional digital payment rate may
	therefore arise from barriers at internet access, basic online
	communication, banking adoption, or payment adoption. Identifying
	where disadvantage emerges is directly relevant for policy design.
	
	\item \textbf{How is digital competence distributed, and which
		dimension explains the gaps that income and health do not?}
	Any residual gap in digital service use after accounting for income,
	health, and preferences must reflect structural barriers --- such as
	connection quality or interface inaccessibility --- or differences in
	digital skill. A principled latent measure of digital literacy is
	needed to assess which component matters more and where skill deficits
	are most concentrated.
\end{enumerate}

\paragraph*{Contributions.}
The four research questions map directly onto four empirical
contributions using the 2020 CIUS, a nationally representative
cross-sectional survey of 17,409 Canadians aged 15 and older
across the ten provinces.
\begin{enumerate}[leftmargin=1.5em, nosep, label=(\roman*)]
	\item We apply the \texttt{svy LLasso} of \citet{JMT2026}, a survey-weighted logistic Lasso designed for variable selection
	in complex survey data, 
	together with debiased post-selection inference
	\citep{Javanmard-Montanari(2014), Zhang-Zhang(2014)}, to
	identify predictors of online banking and digital payment
	adoption from a broad candidate set of sociodemographic
	characteristics, and to measure how these services are
	concentrated across the income distribution.
	\item We decompose age$\,\times\,$education gaps in
	unconditional online banking and health-information use into
	contributions from income, disability, health, and
	preferences using an exact Shapley decomposition
	\citep{Shorrocks2013}.
	\item We estimate a continuation-ratio (sequential) logit
	that models digital adoption as a four-rung ladder: internet
	use, email use, online banking, and digital payments.
	\item We construct a bifactor IRT measure of digital literacy
	\citep{Birnbaum1968, Chalmers2012} and embed it in the
	sequential framework to assess how much of the residual
	education gradient reflects measurable competence.
\end{enumerate}
\paragraph*{Main findings.}
The \texttt{svy LLasso} identifies age, education, and employment as
the most consistent predictors of digital financial service adoption,
with education the only covariate that remains significant at every
rung of the digital ladder. Income-related inequality is significant
across all digital financial services and is especially pronounced for
virtual-wallet adoption; for online banking, employment and education
together account for nearly half of the pro-rich concentration,
indicating a broad socioeconomic gradient rather than a purely
income-based divide.

The age--education decomposition shows that
income, disability, health, and preferences together explain a
substantial but incomplete share of the observed gaps, with
preferences the dominant mediator and large residual gaps persisting
across all cells. The sequential logit shows that disadvantage
emerges at different stages for different groups: education generates
significant barriers throughout the ladder, disability produces its
largest penalty at the digital-payments stage, and the most excluded
profiles --- older, less-educated respondents with low income or
non-employment --- face compounding barriers at multiple rungs.
 
The IRT analysis shows that digital literacy eliminates the education
gradient at internet entry and substantially reduces it at the online
banking rung, but a residual persists, pointing to behavioral and
institutional frictions beyond measurable competence. The IRT
analysis also reveals counterintuitive patterns: a Gen~Z deficit in
structured information-seeking and a security-behavior gap
concentrated among recent immigrants and visible minorities rather
than among seniors.

\paragraph*{Paper organization.}
Section~\ref{sec:data} describes the CIUS 2020 data, and
Section~\ref{sec:framework} sets out the analytical frameworks.
Section~\ref{sec:Q1} presents the \texttt{svy LLasso} results and
concentration-index analysis (Q.1), while Section~\ref{sec:Q2}
reports the age--education decomposition (Q.2).
Section~\ref{sec:Q3} presents the sequential logit analysis (Q.3),
and Section~\ref{sec:IRT} reports the IRT-based digital literacy
measure and mediator analysis (Q.4).
Section~\ref{sec:conc} concludes.
Additional technical details, supplementary results, and robustness
checks are provided in the appendix.
\section{Data}
\label{sec:data}

We use the publicly available 2020 Canadian Internet Use Survey
public-use microdata file (PUMF).\footnote{The dataset is available at
	\url{https://abacus.library.ubc.ca/dataset.xhtml?persistentId=hdl:11272.1/AB2/NUVBX2}.}
The PUMF is close in scope to the Statistics Canada Research Data Centre
version, although some variables are available only in more aggregated
form. CIUS 2020 comprises 17,409 observations on Canadians aged 15 and
older living in one of the ten provinces. First Nations persons living
on reserve are excluded by the sampling frame. The survey uses a
stratified sampling design at the
province/census-metropolitan-area/census-agglomeration level and
combines landline and cellular telephone frames; the overall response
rate is 41.6\%. All estimates use the person weight \texttt{WTPG},
which reflects the adjusted selection probability, non-response
adjustment, and calibration to independent population totals.\footnote{
	Precise variable definitions, item codes, and further information on
	sampling and weighting correspond to the publicly available CIUS 2020
	codebook and User Guide.}

The three main outcomes used in Sections~\ref{sec:Q1}--\ref{sec:Q3}
are internet use, online banking, and digital payments. Internet use is
defined on the full sample. Online banking is defined for internet
users. The digital payments outcome indicates whether the respondent
used a virtual wallet or credit card for online purchases; this outcome
is defined for respondents who ordered goods or services online. The
\texttt{svy LLasso} analysis also considers virtual wallet and credit
card use as separate outcomes; the corresponding results are reported in
Appendix~\ref{app:extra_tables}.\footnote{Email use is not reported as a
	standalone outcome because it is nearly universal among internet users;
	it enters the analysis as the second rung of the sequential logit in
	Section~\ref{sec:Q3}.}

The Q.2 decomposition uses two unconditional outcomes: online banking
and online health-information search. For these outcomes, non-internet
users are coded as zero. \citet{Wavrock_2022} highlight these measures
as especially policy-relevant and well suited to the survey's
sequential routing structure.

The explanatory variables include household income quintile,
educational attainment, employment status, age group, gender,
Indigenous identity, visible minority status, immigration status,
language, household composition, disability status, self-reported
health, preference for non-use, and province. This covariate set is
used across all empirical analyses, with the \texttt{svy LLasso}
specification allowing for a flexible variable-selection step.

\section{Analytical framework}
\label{sec:framework}

\subsection{Survey-weighted logistic Lasso}
\label{sec:svy_lasso}

To address Q.1, we adopt the \texttt{svy LLasso} of
\citet{JMT2026}, which obtains the survey-weighted logistic Lasso estimator
$\hat{\theta}$ by minimizing the penalized weighted negative
log-likelihood:
\begin{equation}\label{eq:svy_lasso}
	\hat{\theta}:=\arg\min_{\theta=(\alpha,\beta')'\in\mathbb{R}^{p+1}}
	\left(-L(\theta)+\lambda\sum_{j=1}^{p}|\beta_j|\right),
\end{equation}
where $\beta=(\beta_1,\dots,\beta_p)'$, 
\begin{equation}\label{eq:logitlik}
	L(\theta):=
	n^{-1}\sum_{i=1}^{n}
	w_i\left(y_i x_i'\theta-\log(1+\exp(x_i'\theta))\right)
\end{equation}
is the weighted log-likelihood of the logit model, and $w_i$ is the survey weight. The intercept $\alpha$ is left unpenalized, and the
$\ell_1$ penalty shrinks weak predictors toward zero. The tuning parameter $\lambda$ is chosen by ten-fold cross-validation
using the AUC criterion, implemented in the \texttt{R} package
\texttt{glmnet}; person weights are rescaled to mean one within the
estimation sample. In Appendix~\ref{app:robustness}, we consider two alternative tuning
rules: a design-aware weighted cross-validation procedure
\citep{Iparragirre2023} and the bootstrap-after-cross-validation rule
of \citet{Chetverikov2025}. Both yield very similar empirical
patterns.

Following \citet{JMT2026}, we apply the debiased Lasso correction to
conduct valid post-selection inference:
\begin{equation}\label{eq:DB}
	\tilde{\theta}^{\mathsf{DB}}
	=
	\hat{\theta}+H(\hat{\theta})^{-1}S(\hat{\theta}),
\end{equation}
where $H(\cdot)$ and $S(\cdot)$ denote the sample Hessian and score of
the full weighted logistic log-likelihood in \eqref{eq:logitlik}. This
one-step estimator removes regularization bias and is asymptotically
normal, facilitating standard $t$-ratio inference on coefficients and
AMEs. Throughout, we report $\tilde{\theta}^{\mathsf{DB}}$ together
with the debiased estimates of the AMEs, denoted by
$\widetilde{\mathrm{AME}}^{\mathsf{DB}}$; technical details are given in
Appendix~\ref{app:inference}.

\subsection{Age--education decomposition framework}
\label{sec:decomp_method}

To address Q.2, we estimate two nested survey-weighted logit models for
each outcome $y_i$:
\begin{align}
	y_i &= \alpha + \theta_g + u_i, \label{eq:m1}\\
	y_i &= \alpha + \theta_g + x_i'\beta + u_i, \label{eq:m2}
\end{align}
where $\theta_g$ denotes the coefficient for the age$\,\times\,$education
cell $g$ relative to the omitted reference cell of adults aged 65 and
older with high school or less education. The control vector $x_i$
includes household-size-adjusted income quintile, a disability
indicator, indicators for self-reported health, and a preference
indicator for non-use due to lack of interest or time.

Let
\begin{equation*}
	m_g(x):=\Lambda(\alpha+\theta_g+x'\beta),
	\qquad
	\Lambda(z)=\frac{e^z}{1+e^z},
\end{equation*}
denote the conditional probability for cell $g$. We partition
the covariate vector into four explanatory blocks,
\begin{equation*}
x=\bigl(x^{(\mathrm{inc})},x^{(\mathrm{dis})},
x^{(\mathrm{health})},x^{(\mathrm{pref})}\bigr),
\end{equation*}
corresponding to income, disability, health, and preferences. We define the characteristics-explained component of the age--education
gap as
\begin{equation}
	E_g:=m_g(x_g)-m_g(x_r),
	\label{eq:Eg_exact}
\end{equation}
where $x_g$ denotes the observed covariate values for respondents in
cell $g$, and $x_r$ denotes the corresponding values for respondents in
the reference cell. Thus, $m_g(x_g)$ is the average fitted probability
evaluated at the observed characteristics of group $g$, whereas
$m_g(x_r)$ evaluates the same fitted probability after replacing those
characteristics with the reference-group values.

To allocate $E_g$ across the four explanatory blocks, we implement the
exact simulation-based Shapley decomposition of \citet{Shorrocks2013},
which adapts the Shapley value \citep{Shapley1953} to regression
decomposition. Let $K$ denote the set of blocks.
For any subset $S \subseteq K$, let
$m_g(x_r^{(-S)},x_g^{(S)})$ denote the fitted probability obtained by
assigning the blocks in $S$ their group-$g$ values and all remaining
blocks their reference-group values. Then the contribution of block $k$
to the explained component for cell $g$ is
\begin{equation}
	C_{gk}
	:=
	\sum_{S \subseteq K \setminus \{k\}}
	\frac{|S|!\,(|K|-|S|-1)!}{|K|!}
	\left[
	m_g\!\left(x_r^{(-S\cup\{k\})},x_g^{(S\cup\{k\})}\right)
	-
	m_g\!\left(x_r^{(-S)},x_g^{(S)}\right)
	\right],
	\label{eq:Cgk_shapley}
\end{equation}
where the bracketed term is the marginal contribution of block $k$
when it is added after the blocks in $S$ have already been switched
from reference-group values to group-$g$ values. By construction, the
decomposition is exact,
\begin{equation*}
	E_g=\sum_{k\in K} C_{gk},
\end{equation*}
and invariant to the ordering of the blocks because it averages
marginal contributions over all possible permutations. These properties
make it especially attractive in the present nonlinear logit setting,
where first-order linearization can perform poorly when fitted
probabilities lie near the boundaries of the unit interval.

We estimate the decomposition on three nested samples --- urban
non-disabled respondents, all urban respondents, and the full sample
--- to distinguish connectivity constraints from disability-related
barriers. The underlying logit models are estimated using
\texttt{svyglm} in the \texttt{R} package \texttt{survey}, and the
Shapley decomposition is then applied to the fitted values.

\subsection{Sequential logit}
\label{sec:seq_method}

The CIUS routing structure described in Section~\ref{sec:data}
implies a natural four-rung adoption ladder: internet use, email use,
online banking, and digital payments. We exploit this structure by
estimating a continuation-ratio logit in which each stage is modelled
as a binary logit on the subsample that has cleared the previous gate.

Let $p_{g,1}$ denote the probability that group $g$ clears Stage~1,
and let $c_{g,j}$ denote the conditional probability of clearing
Stage~$j$ given clearance of Stage~$j-1$. The unconditional probability
of reaching the digital-payments rung is
$\pi_{g,4}=p_{g,1}c_{g,2}c_{g,3}c_{g,4}$. The gap relative to a
benchmark group $\bar g$ admits the exact sequential decomposition
\begin{equation}
	\Delta^{S4}_{g}
	=
	(\bar p_{1}-p_{g,1})\bar c_{2}\bar c_{3}\bar c_{4}
	+ p_{g,1}(\bar c_{2}-c_{g,2})\bar c_{3}\bar c_{4}
	+ p_{g,1}c_{g,2}(\bar c_{3}-c_{g,3})\bar c_{4}
	+ p_{g,1}c_{g,2}c_{g,3}(\bar c_{4}-c_{g,4}),
	\label{eq:seq_decomp}
\end{equation}
obtained by sequentially replacing the benchmark-stage probabilities
with those of group $g$ from Stage~1 onward. The weighting terms
reflect the sequential nature of the process: gaps at later stages
matter only for the subset that clears the earlier rungs. The same
covariate vector is used at each stage, and each stage is estimated
using \texttt{svyglm}.

\subsection{IRT digital literacy score}
\label{sec:irt_method}

To measure digital competence, we estimate a weighted bifactor
two-parameter logistic (2PL) item-response model \citep{Birnbaum1968}
on 20 binary CIUS items covering information-seeking, software and
file management, and security and privacy tasks. Let
$y_{ij}\in\{0,1\}$ denote the response of individual $i$ to item $j$,
where $y_{ij}=1$ if the respondent reports performing the
corresponding digital task and $y_{ij}=0$ otherwise. The model
includes one latent general factor, interpreted as overall digital
literacy, and three latent domain-specific factors that capture
residual covariance within the information-seeking,
software/file-management, and security/privacy domains. The response
probability is specified as
\begin{equation*}
P(y_{ij}=1 \mid \theta_i^{(G)}, \theta_i^{(D_j)})
=
\Lambda\!\left(
a_j^{(G)} \theta_i^{(G)}
+
a_j^{(D_j)} \theta_i^{(D_j)}
-
b_j
\right),
\end{equation*}
where $\Lambda(\cdot)$ is the logistic CDF, $\theta_i^{(G)}$ is the
latent general factor for individual $i$, $\theta_i^{(D_j)}$ is the
latent domain-specific factor for the domain to which item $j$
belongs, $a_j^{(G)}$ and $a_j^{(D_j)}$ are item discrimination
parameters, and $b_j$ is an item difficulty parameter. The latent
factors are assumed to be mutually orthogonal and normalized to unit
variance for identification. The bifactor structure is identified
without rotation because the general factor loads on all items and
each domain-specific factor loads only on items within its domain
\citep{Reise2012}.

Our digital literacy score is the estimated general-factor score
$\hat{\theta}_i^{(G)}$, which summarizes each respondent's overall
digital competence net of domain-specific residual variation. Given the
estimated item parameters
$\hat{\psi}=\{\hat{a}_j^{(G)},\hat{a}_j^{(D_j)},\hat{b}_j\}_{j=1}^{20}$,
this score is computed as the expected a posteriori (EAP) estimate
(see Appendix~\ref{app:EAP} for the formal definition), obtained using
the EM or quasi-Monte Carlo EM algorithm implemented in the
\texttt{R} package \texttt{mirt} \citep{Chalmers2012}, with survey
weights normalized within the estimation sample. For use in the concentration-index 
and sequential-logit analyses, we rescale the estimated general-factor score $\hat{\theta}_i^{(G)}$ to
the unit interval and denote the resulting literacy index by
$\hat{L}_i\in[0,1]$.

The standardized loadings reported in Appendix~\ref{app:irt_loadings} are
obtained via the Schmid--Leiman orthogonalization
\citep{Schmid1957}, which decomposes the factor solution into a
general factor and orthogonal domain-specific residual factors,
ensuring that $\hat{a}_j^{(G)}$ reflects the unique contribution of
the general factor to item $j$ net of domain-specific variance.

Item selection and dimensionality diagnostics are reported in
Appendix~\ref{app:irt}. Reliability is assessed using
McDonald's hierarchical omega \citep{McDonald1999}, which
measures the proportion of composite-score variance attributable
to the general factor net of domain-specific variance; values
above 0.70 are generally taken as evidence that a single general
factor dominates reliable variation.

\subsection{Income-ranked and literacy-ranked concentration indices}
\label{sec:ci_method}

To complement the regression analysis with a distributional
perspective, we use concentration indices to measure whether digital
outcomes are disproportionately concentrated among respondents higher
in the income or digital-literacy distribution.

\paragraph*{Income-ranked concentration index.}
Let $R_i$ denote respondent $i$'s fractional rank in the
household-income distribution, and let $\mu_y=\mathrm{E}[y_i]$ denote
the population mean of outcome $y_i$. The population concentration
index is
\begin{equation*}
	\mathcal{C}_y
	:=
	\frac{2}{\mu_y}\,\mathrm{Cov}(y_i,R_i).
\end{equation*}
A positive value indicates that the outcome is concentrated among
higher-income respondents. In the data, we estimate $\mathcal{C}_y$
using survey weights. With $w_i>0$ denoting the person weight, let
$
W:=\sum_{i=1}^n w_i$
be the total survey weight, and let
\begin{equation*}
\bar{y}
=
\frac{1}{W}\sum_{i=1}^n w_i y_i
\end{equation*}
and $\hat r_i$ denote the weighted sample mean of $y_i$ and the
weighted midpoint fractional rank in the income distribution,
respectively. The survey-weighted sample analogue is
\begin{equation}\label{eq: conc_ind}
	\widehat{\mathcal{C}}_y
	:=
	\frac{2}{\bar{y}\,W}
	\sum_{i=1}^n
	w_i\bigl(y_i-\bar{y}\bigr)\bigl(\hat r_i-\tfrac{1}{2}\bigr).
\end{equation}
Because $y_i$ is directly observed, the influence function of
$\widehat{\mathcal{C}}_y$ admits a closed-form linearization
and inference is based on the resulting first-order variance
estimator \citep{Kakwani1997}.\footnote{In the implementation, we use the equivalent
	three-moment representation
	$\widehat{\mathcal{C}}_y
	=
	2\bar{y}^{-1}W^{-1}\sum_i w_i y_i \hat{r}_i
	-
	1
	-
	2(W^{-1}\sum_i w_i \hat{r}_i - \tfrac{1}{2})$,
	which reduces to the formula \eqref{eq: conc_ind} when
	$W^{-1}\sum_i w_i \hat{r}_i = \tfrac{1}{2}$ exactly.
	After excluding observations with missing outcome values,
	the weighted mean rank in the estimation sample need not
	equal $\tfrac{1}{2}$, so the three-moment form is used
	throughout.}

Since the outcomes of interest are binary, the standard
concentration index is bounded by $[-(1-\mu_y),\,1-\mu_y]$ rather than
$[-1,1]$, which complicates comparisons across outcomes with different
means \citep{Wagstaff2005}. We therefore also report two normalized versions. Following
\citet{Wagstaff2005}, the Wagstaff-normalized index is
\begin{equation*}
	\widehat{\mathcal{C}}_y^{\mathrm{Wag}}
	=
	\frac{\widehat{\mathcal{C}}_y}{1-\bar{y}},
\end{equation*}
and, following \citet{Erreygers2009}, the Erreygers index is
\begin{equation*}
	\widehat{\mathcal{C}}_y^{\mathrm{E}}
	=
	4\bar{y}\,\widehat{\mathcal{C}}_y.
\end{equation*}
The income-ranked decomposition in Section~\ref{sec:CI} uses the
standard index, which admits the standard linear decomposition of
\citet{Wagstaff2003}.

\paragraph*{Literacy-ranked concentration index.}
To study how digital activities are distributed across the competence
distribution, we define a literacy-ranked concentration index by
replacing the income rank with the weighted rank of the estimated
general-factor digital literacy score. Let $r_i^L$ denote respondent
$i$'s weighted midpoint fractional rank in that literacy distribution. 
The corresponding population index is
\begin{equation*}
	\mathcal{C}_y^L
	:=
	\frac{2}{\mu_y}\,\mathrm{Cov}(y_i,r_i^L),
\end{equation*}
with survey-weighted sample analogue
\begin{equation*}
	\widehat{\mathcal{C}}_y^L
	:=
	\frac{2}{\bar{y}\,W}
	\sum_{i=1}^n
	w_i\bigl(y_i-\bar{y}\bigr)\bigl(\hat r_i^L-\tfrac{1}{2}\bigr).
\end{equation*}
A positive value indicates that the outcome is concentrated among the
more digitally literate.

Because the literacy rank is constructed from an estimated bifactor
IRT model, $\widehat{\mathcal{C}}_y^L$ is a two-step estimator.
Inference therefore uses Murphy--Topel-type standard errors \citep{MurphyTopel1985} that account
for both second-stage sampling variation and first-stage IRT
estimation uncertainty; details are in
Appendix~\ref{app:variance}.

\section{Q.1: Who is excluded? Predictors of digital engagement}
\label{sec:Q1}
	
\subsection{Online banking}

Table~\ref{tab:lasso_bank_pumf} presents the \texttt{svy LLasso}
results for the online banking model. Age is the strongest predictor:
relative to the reference group aged 45--54, respondents aged 25--34
and 35--44 are 7 and 6 percentage points more likely to use online
banking, while those aged 55--64 and 65 and older are 3 and 9
percentage points less likely. Employment status is also important:
employed respondents are about 9 percentage points more likely to use
online banking ($\widetilde{\mathrm{AME}}^{\mathsf{DB}}=0.09$,
$p<0.001$), consistent with payroll direct deposit and
employer-linked financial access. Educational attainment generates a
clear gradient: high school (HS) or less reduces the probability of online
banking by 7 percentage points, while a university degree raises it by
5 percentage points relative to some post-secondary. Visible minority
status is associated with a 5 percentage point lower probability
($p=0.003$). Among provinces, Manitoba is associated with a 5
percentage point lower probability ($p=0.010$) and Quebec with a 5
percentage point higher probability ($p=0.021$). Gender, rural
residence, Aboriginal identity, immigration status, and disability are
not statistically significant after conditioning on the full covariate
set.

Language of use is also a notable predictor. Relative to respondents
reporting neither English nor French, English speakers and
English-French speakers are 17 and 13 percentage points more likely
to use online banking, respectively. These are among the largest AMEs
in the model, though their magnitude partly reflects the small and
linguistically heterogeneous reference category rather than a uniform
language penalty. Within the Canadian financial system, online banking
interfaces have historically been designed around English and French,
and some platforms impose official language requirements for digital
enrollment. The language gradient is therefore consistent with a
combination of interface accessibility barriers and the occupational
and socioeconomic sorting that correlates with language of use. The
French-speaker coefficient is positive but falls short of conventional
significance ($p=0.178$), consistent with Quebec's above-average
online banking rate.

\renewcommand{\arraystretch}{0.75}
\newcommand{\midsize}{\fontsize{9pt}{10pt}\selectfont}
\begin{table}[htbp]
	\small
	\begin{threeparttable}
		\caption{Lasso Logistic Regression Results: Online Banking}
		\label{tab:lasso_bank_pumf}
		\begin{tabular}{llLLrLr}
			\toprule
			\multicolumn{1}{l}{Variables} &
			\multicolumn{1}{l}{Categories} &
			\multicolumn{1}{l}{$\texttt{svy LLasso}$} &
			\multicolumn{1}{l}{$\tilde{\theta}^{\mathsf{DB}}$} &
			\multicolumn{1}{l}{p-value} &
			\multicolumn{1}{l}{$\widetilde{\mathrm{AME}}^{\mathsf{DB}}$} &
			\multicolumn{1}{l}{p-value} \\
			\midrule
			\textit{Intercept} &  & 1.09 & -0.15 & 0.757 & \multicolumn{1}{c}{-} & - \\
			\textit{Location} & Rural & \multicolumn{1}{c}{-} & -0.04 & 0.583 & -0.01 & 0.591 \\
			\textit{Age} & 15--24 & -0.00 & -0.21 & 0.132 & -0.03 & 0.129 \\
			& 25--34 & 0.42 & 0.58^{***} & $<0.001$ & 0.07^{***} & $<0.001$ \\
			& 35--44 & 0.32 & 0.51^{***} & $<0.001$ & 0.06^{***} & $<0.001$ \\
			& 55--64 & \multicolumn{1}{c}{-} & -0.23^{**} & 0.024 & -0.03^{**} & 0.023 \\
			& 65 and older & -0.34 & -0.60^{***} & $<0.001$ & -0.09^{***} & $<0.001$ \\
			\textit{Gender} & Female & \multicolumn{1}{c}{-} & 0.09 & 0.162 & 0.01 & 0.173 \\
			\textit{Aboriginal identity} & Aboriginal & \multicolumn{1}{c}{-} & -0.16 & 0.381 & -0.02 & 0.377 \\
			\textit{Language} & English & \multicolumn{1}{c}{-} & 1.13^{**} & 0.014 & 0.17^{**} & 0.010 \\
			& French & \multicolumn{1}{c}{-} & 0.80^{*} & 0.095 & 0.09 & 0.178 \\
			& Eng and Fr & 0.02 & 1.14^{**} & 0.015 & 0.13^{**} & 0.045 \\
			\textit{Employment} & Employed & 0.68 & 0.62^{***} & $<0.001$ & 0.09^{***} & $<0.001$ \\
			\textit{Education} & HS or less & -0.50 & -0.47^{***} & $<0.001$ & -0.07^{***} & $<0.001$ \\
			& University degree & 0.29 & 0.35^{***} & $<0.001$ & 0.05^{***} & $<0.001$ \\
			\textit{Minority} & Visible minority & -0.16 & -0.32^{***} & 0.002 & -0.05^{***} & 0.003 \\
			\textit{Household type} & Family w/o child & 0.13 & 0.35^{***} & $<0.001$ & 0.05^{***} & $<0.001$ \\
			& Single & \multicolumn{1}{c}{-} & 0.08 & 0.414 & 0.01 & 0.434 \\
			& Other household & \multicolumn{1}{c}{-} & 0.38^{*} & 0.067 & 0.05 & 0.105 \\
			\textit{Income} & Income Q1 & \multicolumn{1}{c}{-} & -0.01 & 0.949 & 0.00 & 0.951 \\
			& Income Q3 & \multicolumn{1}{c}{-} & 0.15 & 0.134 & 0.02 & 0.154 \\
			& Income Q4 & \multicolumn{1}{c}{-} & 0.24^{**} & 0.019 & 0.03^{**} & 0.028 \\
			& Income Q5 & \multicolumn{1}{c}{-} & 0.26^{**} & 0.015 & 0.03^{**} & 0.023 \\
			\textit{Immigration} & Non-landed & \multicolumn{1}{c}{-} & -0.03 & 0.828 & 0.00 & 0.833 \\
			\textit{Disability} & Disabled & \multicolumn{1}{c}{-} & -0.16 & 0.190 & -0.02 & 0.188 \\
			\textit{General health} & Excellent & \multicolumn{1}{c}{-} & -0.10 & 0.279 & -0.01 & 0.286 \\
			& Very good & 0.13 & 0.21^{***} & 0.008 & 0.03^{***} & 0.009 \\
			& Fair & \multicolumn{1}{c}{-} & 0.21^{*} & 0.076 & 0.03^{*} & 0.099 \\
			& Poor & \multicolumn{1}{c}{-} & 0.11 & 0.595 & 0.01 & 0.615 \\
			\textit{Province} & NL & \multicolumn{1}{c}{-} & 0.05 & 0.752 & 0.01 & 0.761 \\
			& PEI & \multicolumn{1}{c}{-} & -0.01 & 0.956 & 0.00 & 0.957 \\
			& NS & \multicolumn{1}{c}{-} & -0.03 & 0.815 & 0.00 & 0.819 \\
			& NB & \multicolumn{1}{c}{-} & -0.03 & 0.822 & 0.00 & 0.826 \\
			& QC & 0.11 & 0.36^{**} & 0.015 & 0.05^{**} & 0.021 \\
			& ON & \multicolumn{1}{c}{-} & -0.05 & 0.650 & -0.01 & 0.658 \\
			& MB & \multicolumn{1}{c}{-} & -0.36^{**} & 0.015 & -0.05^{**} & 0.010 \\
			& SK & \multicolumn{1}{c}{-} & -0.05 & 0.756 & -0.01 & 0.760 \\
			& BC & \multicolumn{1}{c}{-} & -0.03 & 0.810 & 0.00 & 0.814 \\
			\bottomrule
		\end{tabular}
		\begin{tablenotes}
		\midsize{\item \textit{Notes}: $n = 15,020$. $\tilde{\theta}^{\mathsf{DB}}$ and $\widetilde{\mathrm{AME}}^{\mathsf{DB}}$ denote debiased Lasso estimates of the logit parameter and AME respectively. ``-'' denotes variables not selected by \texttt{svy LLasso}. Comparison categories and not-stated responses are omitted for brevity. Reference: urban, age 45--54, male, non-Aboriginal, neither English nor French, not employed, some post-secondary, non-visible minority, family with child, income Q2, landed immigrant, not disabled, omitted health category, Alberta. Significance levels: *** $p<0.01$, ** $p<0.05$, * $p<0.10$. Exact p-values are reported; values below 0.001 are shown as $<0.001$.}
		\end{tablenotes}
	\end{threeparttable}
\end{table}

\subsection{Digital payments}\label{sec:digital_payments}

Table~\ref{tab:Logit_vrwallet_credit_pumf} presents the
\texttt{svy LLasso} results for virtual wallet and credit card use.
For virtual wallet adoption, age remains the dominant predictor:
those aged 15--24, 25--34, and 35--44 are 11, 8, and 4 percentage
points more likely to adopt relative to the reference group, while
those aged 55--64 and 65 and older are 6 and 8 points less likely.
Rural residence reduces adoption by 5 points. Unlike in the
online-banking model, visible minority status has a positive
association ($\widetilde{\mathrm{AME}}^{\mathsf{DB}}\approx 0.04$, $p<0.01$), pointing to a distinct
adoption pattern rather than a general financial-inclusion gradient.
University education and top-quintile income are also positively
associated, with AMEs of about 2 and 8 percentage points.

For credit card use, the youngest group (15--24) is about 9
percentage points less likely to use a credit card online. Education
generates the sharpest gradient: high school or less reduces the
probability by 7 percentage points, while a university degree raises
it by 7 points. Family households without children and Ontario
residence are positively associated, while disability is associated
with a 7 percentage point reduction and Quebec with a 6 point
reduction. Visible minority status is weakly negative for credit
card use, in contrast to its positive association with virtual wallet
adoption. This sign reversal is consistent with substitution between
payment instruments and differential access barriers across
socioeconomic groups.

\renewcommand{\arraystretch}{0.78}
\begin{table}[ht]
	\small
	\begin{threeparttable}
		\caption{Lasso Logistic Regression Results: Digital Payments}
		\label{tab:Logit_vrwallet_credit_pumf}
		\begin{tabular}{llLLLLLL}
			\toprule
			& &
			\multicolumn{3}{c}{Virtual wallet} &
			\multicolumn{3}{c}{Credit card} \\
			\cmidrule(lr){3-5}\cmidrule(lr){6-8}
			Variables & Categories &
			\multicolumn{1}{l}{$\texttt{svy LLasso}$} &
			\multicolumn{1}{l}{$\tilde{\theta}^{\mathsf{DB}}$} &
			\multicolumn{1}{l}{$\widetilde{\mathrm{AME}}^{\mathsf{DB}}$} &
			\multicolumn{1}{l}{$\texttt{svy LLasso}$} &
			\multicolumn{1}{l}{$\tilde{\theta}^{\mathsf{DB}}$} &
			\multicolumn{1}{l}{$\widetilde{\mathrm{AME}}^{\mathsf{DB}}$} \\
			\midrule
			\textit{Location} & Rural
			& -0.16 & -0.56^{***} & -0.05^{***}
			& -     & 0.07         & 0.01 \\
			
			\textit{Age} & 15--24
			& 0.30  & 0.83^{***}  & 0.11^{***}
			& -0.41 & -0.54^{***} & -0.09^{***} \\
			& 25--34
			& 0.22  & 0.63^{***}  & 0.08^{***}
			& -     & 0.04         & 0.01 \\
			& 35--44
			& -     & 0.36^{***}  & 0.04^{***}
			& -     & 0.12         & 0.02 \\
			& 55--64
			& -0.31 & -0.59^{***} & -0.06^{***}
			& -     & -0.03        & 0.00 \\
			& 65 and older
			& -0.54 & -0.96^{***} & -0.08^{***}
			& -     & -0.09        & -0.01 \\
			
			\textit{Education} & HS or less
			& -     & -0.04        & 0.00
			& -0.42 & -0.43^{***} & -0.07^{***} \\
			& Univ. degree
			& 0.03  & 0.21^{**}   & 0.02^{**}
			& 0.39  & 0.49^{***}  & 0.07^{***} \\
			
			\textit{Minority} & Visible min.
			& 0.18  & 0.37^{***}  & 0.04^{***}
			& -0.02 & -0.19^{*}   & -0.03^{*} \\
			
			\textit{Household} & Fam. w/o child
			& -     & 0.08         & 0.01
			& 0.09  & 0.34^{***}   & 0.05^{***} \\
			
			\textit{Income} & Income Q1
			& -     & 0.19         & 0.02
			& -0.08 & -0.21^{*}    & -0.03^{*} \\
			& Income Q4
			& -     & 0.23^{*}     & 0.03^{*}
			& -     & 0.19^{*}     & 0.03^{*} \\
			& Income Q5
			& 0.25  & 0.68^{***}   & 0.08^{***}
			& -     & 0.13         & 0.02 \\
			
			\textit{Immigration} & Non-landed
			& -     & 0.26^{*}     & 0.03^{*}
			& -     & 0.16         & 0.02 \\
			
			\textit{Disability} & Disabled
			& -     & -0.01        & 0.00
			& -     & -0.44^{***}  & -0.07^{***} \\
			
			\textit{General health} & Excellent
			& -     & 0.22^{*}     & 0.03^{*}
			& -     & -0.21^{**}   & -0.03^{**} \\
			
			\textit{Province} & MB
			& -     & -0.44^{**}   & -0.04^{**}
			& -     & 0.03         & 0.00 \\
			& ON
			& -     & 0.01         & 0.00
			& 0.06  & 0.24^{**}    & 0.04^{**} \\
			& QC
			& -     & -0.11        & -0.01
			& -0.42 & -0.38^{**}   & -0.06^{**} \\
			\bottomrule
		\end{tabular}
		\begin{tablenotes}
			\footnotesize{\item \textit{Notes}: $n = 12{,}124$ for both models.
				Condensed PUMF version based on the full virtual-wallet and credit-card tables.
				Comparison categories and not-stated responses are omitted for brevity.
				Reference: urban, age 45--54, male, non-Aboriginal, neither English nor French,
				not employed, some post-secondary, non-visible minority, family with child,
				income Q2, landed immigrant, not disabled, health category 3, Alberta.
				Significance levels: *** $p<0.01$, ** $p<0.05$, * $p<0.10$.}
		\end{tablenotes}
	\end{threeparttable}
\end{table}
	
\subsection{Concentration of digital banking access across income}
\label{sec:CI}

Table~\ref{tab:ci_combined} reports the
standard concentration index together with the Wagstaff-normalized and
Erreygers versions for online banking, virtual wallet use, credit card
use, and the composite digital-payments outcome. Inference for the
standard concentration index is based on a first-order linearization
estimator.

Online banking is significantly
concentrated among higher-income respondents:
$\widehat{\mathcal{C}}_y=0.027$ (SE $=0.003$; 95\% CI $[0.020,\,0.034]$;
$p<0.001$). Panel~B shows that the \citet{Wagstaff2003} decomposition
associates 33.1\% of this concentration with income, 26.6\% with
employment, 22.3\% with education, and 14.4\% with age. Employment and
education together account for 48.9\% --- nearly one half --- while
income, although the single largest component, does not dominate the
decomposition on its own. Inequality in digital banking access
therefore reflects a broad socioeconomic gradient rather than a purely
income-based divide.

The comparison across digital-finance outcomes in Panel~A reveals a
much sharper income gradient for newer payment technologies.
Virtual-wallet use is far more concentrated among higher-income
respondents ($\widehat{\mathcal{C}}_y=0.106$, SE $=0.023$, 95\% CI
$[0.061,\,0.152]$; $p<0.001$) than online banking ($0.027$), credit
card use ($0.023$), or the broader digital-payments measure ($0.024$).
The grouped decompositions in Panel~B show that income accounts for
93.0\% of the measured concentration in virtual-wallet use, compared
with 51.6\% for credit card use and 57.8\% for the composite
digital-payments measure. The role of income therefore becomes much
more prominent as one moves from general digital banking toward newer
payment instruments.
\begin{table}[ht]
	\begin{center}
	\small
		\caption{Income-Related Concentration of Digital Financial Services}
		\label{tab:ci_combined}
		\begin{tabular}{lcccc}
			\toprule
			\multicolumn{5}{l}{{Panel A. Concentration Indices}} \\
			\midrule
			Outcome &
			Mean &
			$\widehat{\mathcal{C}}_y$ &
			$\widehat{\mathcal{C}}_y^{\mathrm{Wag}}$ &
			$\widehat{\mathcal{C}}_y^{\mathrm{E}}$ \\
			\midrule
			Online banking & 0.817 & 0.027 & 0.147 & 0.088 \\
			Virtual wallet & 0.123 & 0.106 & 0.121 & 0.053 \\
			Credit card & 0.793 & 0.023 & 0.110 & 0.072 \\
			Digital payments (composite) & 0.812 & 0.024 & 0.129 & 0.079 \\
			\addlinespace
			\midrule
			\multicolumn{5}{l}{{Panel B. Grouped Decomposition Shares (\%)}} \\
			\midrule
			Outcome &
			Income &
			Employment &
			Education &
			Age \\
			\midrule
			Online banking & 33.1 & 26.6 & 22.3 & 14.4 \\
			Virtual wallet & 93.0 & -1.3 & 8.0 & -1.3 \\
			Credit card & 51.6 & 4.0 & 30.5 & 3.0 \\
			Digital payments (composite) & 57.8 & 2.5 & 24.9 & 2.4 \\
			\bottomrule
		\end{tabular}
\end{center}
			\footnotesize 
			\textit{Notes}: Panel A reports the standard concentration index,
			$\widehat{\mathcal{C}}_y$, together with the Wagstaff-normalized and
			Erreygers versions for each digital financial service. Positive
			values indicate a pro-rich gradient. Panel B reports grouped
			percentage contributions from the Wagstaff-type decomposition of the
			standard concentration index. Positive values indicate that the
			covariate group contributes to the pro-rich gradient; negative
			values indicate an offsetting contribution. Smaller grouped
			contributions are omitted for brevity.
\end{table}
\section{Q.2: Why do age--education gaps persist?}
\label{sec:Q2}

This section focuses on \emph{unconditional} digital use. Unlike
conditional measures, unconditional outcomes assign zero to those
who do not reach a given stage, thereby capturing cumulative
barriers along the full pathway to digital participation.
Unadjusted rates and gaps relative to the reference cell are
reported in Appendix~\ref{app:decomp_robustness}; for online
banking, rates range from 0.381 for the reference cell to 0.977
for the 15--24 $\times$ University cell, an unadjusted gap of
0.596.

\subsection{Adjusted gaps and decomposition}

Table~\ref{tab:decomp_adj} reports the exact Shapley decomposition
of age--education gaps after conditioning on income, disability,
health, and preferences; results for health-information search are
included alongside online banking as a cross-outcome robustness
check and show qualitatively similar patterns throughout.
Adjustment reduces the gaps but does not eliminate them. In the
full sample, the adjusted gap in online banking remains 0.379 for
the 15--24 $\times$ University cell, 0.261 for the 25--44
$\times$ University cell, and 0.225 for the 45--64 $\times$
University cell.

Among the mediators, preferences account for the largest explained
share in most cells. Income contributes a meaningful but smaller
portion. Disability and health effects are generally modest; health
contributions are sometimes slightly negative, indicating that
differences in self-reported health do not reinforce the
age--education gradient once other covariates are held fixed.

The total explained component is substantial but incomplete: it is
0.202 for the 15--24 $\times$ University cell and 0.247 for the
25--44 $\times$ University cell in online banking, leaving sizable
residual gaps. These persistent adjusted gaps point to digital skills,
confidence, interface familiarity, and other unobserved barriers.
Subsample results for urban and urban non-disabled respondents are
qualitatively similar and are reported in
Appendix~\ref{app:decomp_robustness}. Section~\ref{sec:IRT} uses the
IRT literacy score to assess how much of the residual is accounted for
by measurable digital competence.

\begin{table}[h!]
	\centering
	\caption{Exact Shapley Decomposition of Age--Education Gaps in Unconditional Digital Use}
	\label{tab:decomp_adj}
	\begin{threeparttable}
		\setlength{\tabcolsep}{4pt}
		
		\begin{tabular}{lrrrrrrr}
			\toprule
			Cell & Obs. gap & Adj. gap & Income & Disability & Health & Preferences & Total expl. \\
			\midrule
			\multicolumn{8}{l}{\textit{Online banking}} \\
			15--24 $\times$ HS or less   & 0.257 & 0.054 & 0.000 & -0.001 & 0.002 & 0.189 & 0.190 \\
			15--24 $\times$ Some PSE     & 0.436 & 0.189 & 0.014 & 0.002  & -0.003 & 0.232 & 0.247 \\
			15--24 $\times$ University   & 0.596 & 0.379 & 0.011 & 0.001  & 0.000  & 0.190 & 0.202 \\
			25--44 $\times$ HS or less   & 0.448 & 0.222 & 0.001 & -0.002 & 0.004  & 0.226 & 0.231 \\
			25--44 $\times$ Some PSE     & 0.521 & 0.277 & 0.003 & 0.001  & 0.000  & 0.239 & 0.242 \\
			25--44 $\times$ University   & 0.535 & 0.261 & 0.011 & 0.002  & 0.003  & 0.231 & 0.247 \\
			45--64 $\times$ HS or less   & 0.260 & 0.100 & 0.013 & -0.008 & -0.001 & 0.146 & 0.152 \\
			45--64 $\times$ Some PSE     & 0.424 & 0.175 & 0.013 & 0.007  & 0.008  & 0.219 & 0.247 \\
			45--64 $\times$ University   & 0.491 & 0.225 & 0.015 & 0.001  & -0.003 & 0.240 & 0.254 \\
			65-plus $\times$ Some PSE    & 0.203 & 0.046 & 0.016 & 0.005  & -0.001 & 0.144 & 0.163 \\
			65-plus $\times$ University  & 0.329 & 0.099 & 0.024 & -0.004 & -0.005 & 0.181 & 0.196 \\
			
			\addlinespace
			\multicolumn{8}{l}{\textit{Health-information search}} \\
			15--24 $\times$ HS or less   & 0.345 & 0.119 & 0.038 & 0.000  & -0.013 & 0.198 & 0.223 \\
			15--24 $\times$ Some PSE     & 0.330 & 0.110 & 0.033 & -0.002 & -0.016 & 0.198 & 0.213 \\
			15--24 $\times$ University   & 0.506 & 0.246 & 0.014 & -0.002 & -0.011 & 0.241 & 0.243 \\
			25--44 $\times$ HS or less   & 0.370 & 0.150 & 0.022 & -0.001 & -0.004 & 0.203 & 0.218 \\
			25--44 $\times$ Some PSE     & 0.447 & 0.191 & 0.028 & 0.002  & -0.006 & 0.224 & 0.248 \\
			25--44 $\times$ University   & 0.531 & 0.281 & 0.018 & -0.003 & -0.006 & 0.224 & 0.233 \\
			45--64 $\times$ HS or less   & 0.186 & 0.034 & 0.030 & 0.000  & -0.002 & 0.123 & 0.151 \\
			45--64 $\times$ Some PSE     & 0.347 & 0.116 & 0.030 & 0.003  & -0.008 & 0.196 & 0.220 \\
			45--64 $\times$ University   & 0.475 & 0.212 & 0.034 & 0.001  & -0.008 & 0.230 & 0.257 \\
			65-plus $\times$ Some PSE    & 0.242 & 0.079 & 0.015 & 0.005  & -0.003 & 0.148 & 0.166 \\
			65-plus $\times$ University  & 0.379 & 0.174 & 0.023 & -0.001 & -0.010 & 0.191 & 0.204 \\
			\bottomrule
		\end{tabular}
		
		\begin{tablenotes}
			\footnotesize
			\item \textit{Notes}: The reference cell is individuals aged 65 and older with high school education or less. ``Obs. gap'' is the weighted difference in the unconditional probability relative to the reference cell. ``Adj. gap'' is the remaining gap after replacing the mediator distribution with that of the reference cell. Mediator columns report exact Shapley contributions. ``Total expl.'' is the exact explained component; mediator contributions sum to this value up to machine precision.
		\end{tablenotes}
		
	\end{threeparttable}
\end{table}

\section{Q.3: Where do groups fall off the digital ladder?}
\label{sec:Q3}

\subsection{Stage-specific marginal effects and model fit}

\begin{figure}[htbp]
	\caption{Stage-Specific Average Marginal Effects from the Sequential Logit}
	\label{fig:ame_heatmap}
	\includegraphics[width=1.1\linewidth, height=1.05\textheight,
	keepaspectratio]{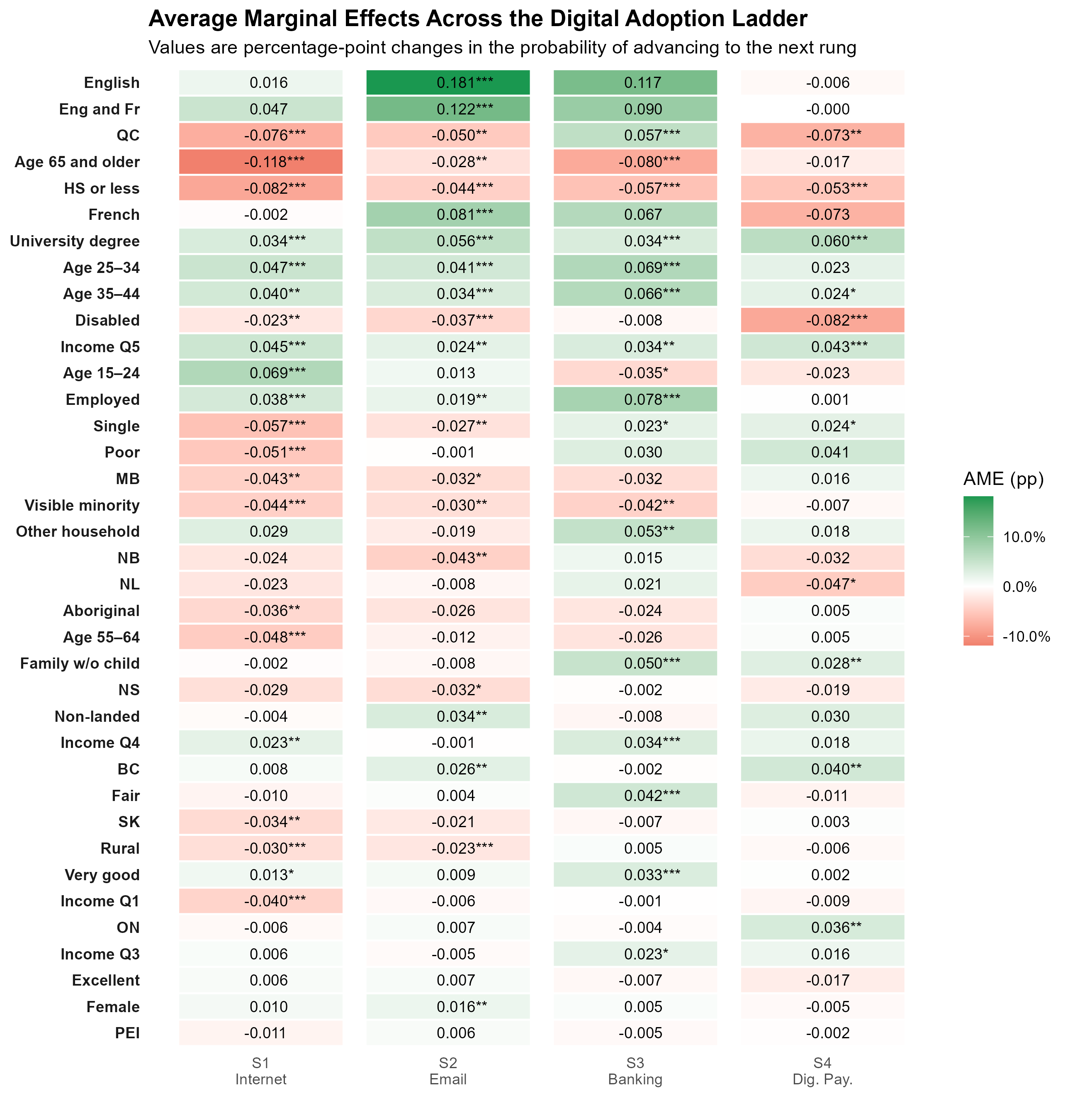}
	\footnotesize{
		\textit{Notes}: Cells report AMEs from the four sequential survey-weighted logit models. Values are expressed in percentage-point changes in the probability of advancing to the next rung. Categories corresponding to ``not stated'' responses are not shown. Reference categories: urban residence, age 45--54, male, non-Indigenous, neither English nor French, not employed, some post-secondary education, non-visible minority, family household with children under 18, income quintile~2, landed immigrant, no disability, health category 3, Alberta. Significance levels: *** $p<0.01$, ** $p<0.05$, * $p<0.10$.
	}
\end{figure}

Figure~\ref{fig:ame_heatmap} reports average marginal effects (AMEs)
from the four sequential logit models. The sequential logit is
estimated as a continuation-ratio model with one binary logit at each
rung of the adoption ladder. Model fit declines monotonically across
stages: McFadden's pseudo-$R^2$ falls from 0.29 at internet entry to
0.12 for email use, 0.09 for online banking, and 0.06 for digital
payments. Sociodemographic characteristics therefore explain a
substantial share of variation in \emph{who gets online}, but
progressively less of the variation in higher-order activities. This
pattern is consistent with early barriers reflecting structural
constraints, while later stages depend more on trust, preferences, and
transaction-specific behavior. Full goodness-of-fit statistics are
reported in Appendix~\ref{app:seq_gof}.

The ladder is characterized by different barriers at different stages.
Low educational attainment is the only disadvantage that remains
significant throughout the entire sequence, from internet access to
digital payment adoption, suggesting that education captures not only
access differences but also the skills and confidence needed for
progressively more demanding digital tasks.

Income-related gaps become sharper at later stages, with the role of
financial capacity most pronounced for digital-payment adoption. Rural
disadvantage shows the opposite pattern: it is concentrated at the
early stages --- especially internet and email use --- and largely
disappears once earlier access barriers are cleared. Gender is largely
neutral across the ladder, with one exception: female respondents are
1.6 percentage points more likely to use email conditional on internet
access ($p = 0.022$), consistent with occupational sorting toward
communication-intensive roles.

Age-related gaps are largest at initial entry and narrow at later
stages, suggesting that older adults face stronger barriers to getting
online than to using digital transactions conditional on access.
Disability, by contrast, is associated with negative effects across
multiple stages, consistent with barriers that persist beyond entry.
Figure~\ref{fig:archetypes} in
Appendix~\ref{app:seq_logit} translates these stage-specific marginal
effects into predicted adoption trajectories for six representative
profiles, illustrating how multiple disadvantages compound across the
ladder.

\subsection{Cumulative reach and drop-off patterns}

\begin{figure}[h!]
	\caption{Survival Curves: Cumulative Reach at Each Rung}
	\includegraphics[width=1\linewidth,height=1\textheight,
	keepaspectratio]{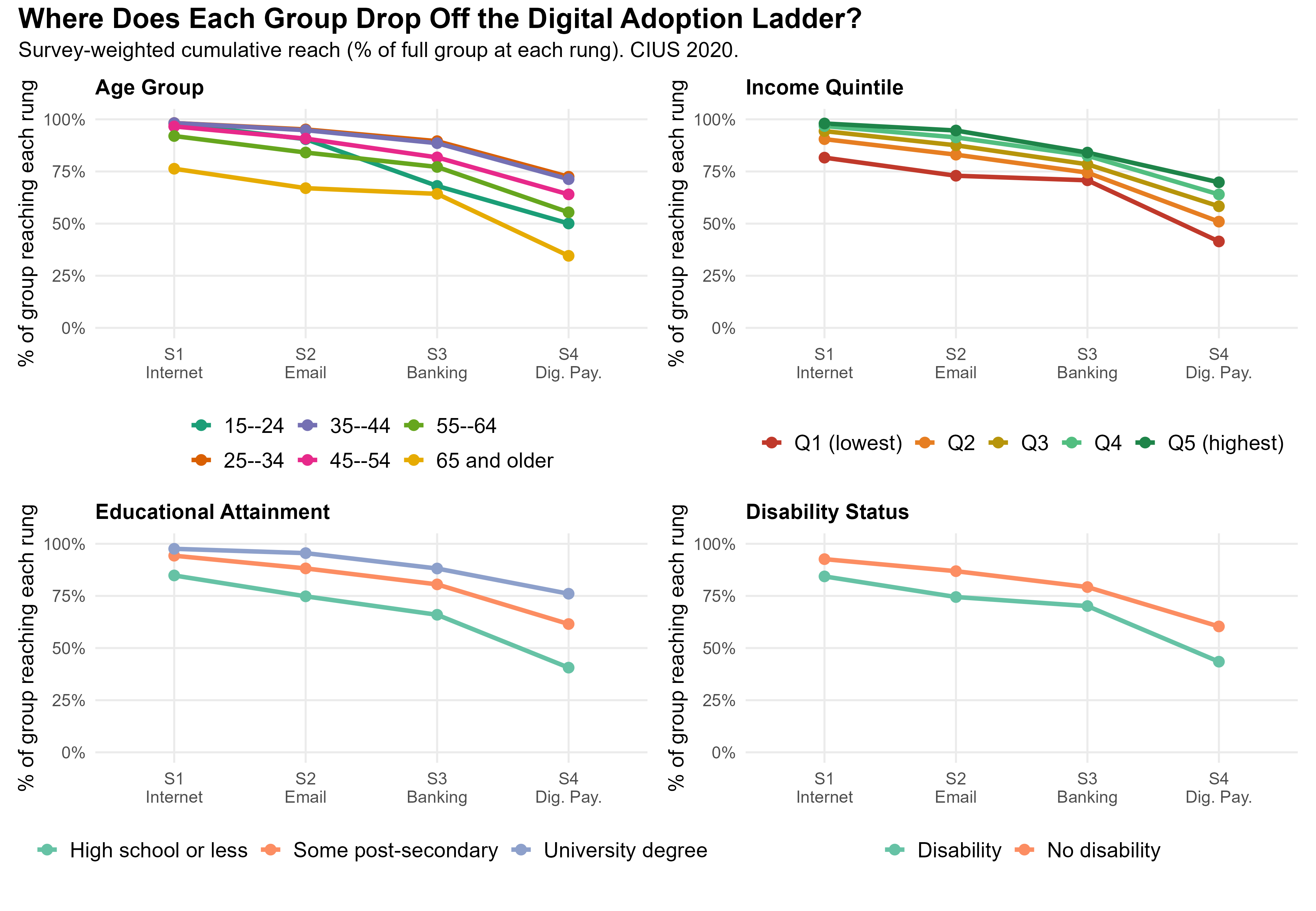}
	\label{fig:ladder_survival}
	\footnotesize{\textit{Notes}: Survey-weighted cumulative reach at each rung by demographic
		group. The panels report cumulative reach for visible minority status,
		income quintile, educational attainment, and disability status.}
\end{figure}

Figure~\ref{fig:ladder_survival} traces the cumulative share of each
demographic group reaching each rung.

\paragraph*{Income inequality becomes most visible at the digital payments rung.}
Income quintile groups remain close through Stages~1--3 but diverge
sharply at Stage~4: roughly 42\% of respondents in Income~Q1
reach digital payments, compared with about 70\% in Income~Q5.
This gap reflects accumulated advantages rather than a single
transition: Income~Q5 raises the probability of internet use by 4.5
points, email by 2.4, online banking by 3.4, and digital payments by
4.3. Income~Q1, by contrast, reduces internet adoption by 4.0 points
while showing little additional conditional disadvantage at later
stages.

\paragraph*{Education is the one disadvantage that persists at every rung.}
High school or less generates a significant negative AME at all four
stages: $-0.082$, $-0.044$, $-0.057$, and $-0.053$. A university
degree produces significant positive increments at every rung:
$0.034$, $0.056$, $0.034$, and $0.060$. No other covariate
produces significant effects uniformly across all four stages. The
persistence of education --- even after controlling for income, age,
employment, disability, and health --- motivates the IRT mediator
analysis in Section~\ref{sec:IRT}.

\paragraph*{Disability generates compounding but non-monotonic disadvantage.}
The AMEs are $-0.023$ at Stage~1, $-0.037$ at Stage~2, $-0.008$ at
Stage~3 (not significant), and $-0.082$ at Stage~4. The pattern is
cumulative but not monotonic. Online banking interfaces appear
relatively more standardized for persons with disabilities, whereas
retail payment checkout flows remain difficult to navigate. This points
to the importance of accessibility enforcement beyond core banking
platforms.

\paragraph*{Employment is specific to the banking rung.}
Employment shows its largest AME at Stage~3 ($0.078$, $p<0.01$),
while its effect at Stage~4 is essentially zero. The mechanisms
connecting employment to banking --- payroll direct deposit, benefits
management, employer-provided financial portals --- appear specific to
the banking rung and do not carry through to retail payment adoption.

\paragraph*{Age effects and geographic heterogeneity.}
Respondents aged 65 and older are 11.8 percentage points less likely
to use the internet, 2.8 points less likely to use email, and
8.0 points less likely to use online banking, with no significant
difference at digital payments. The 15--24 group shows a distinct
profile: substantially more likely to use the internet ($0.069$) but
less likely to use online banking ($-0.035$). Age therefore operates
primarily through entry and banking barriers rather than a uniform
resistance to all digital activities.

Rural residence reduces internet use by 3.0 points and email use by
2.3 points but has no meaningful effect at later stages, indicating
that rural disadvantage in the PUMF data operates mainly through
initial connectivity. Quebec displays a distinctive profile: negative
effects at Stages~1--2, a positive banking effect ($0.057$), and a
negative digital payments effect ($-0.073$), consistent with strong
banking-platform penetration alongside lower adoption of newer payment
methods.

\subsection{Empirically grounded disadvantaged profiles}
\label{sec:disadvantaged_profiles}

To identify which combinations of characteristics jointly
characterize the bottom tail of predicted digital payment
adoption, we examine the modal characteristics among
respondents in the bottom decile of the predicted adoption
distribution, retaining those whose weighted modal share exceeds
0.80 (0.75 for virtual wallet). Appendix~\ref{app:profiles}
reports the corresponding profile summaries and predicted
probabilities.

The virtual wallet exclusion profile is especially stark:
respondents aged 65 and older, with high school or less
education, in the lowest income quintile, and with non-visible
minority status have a predicted adoption probability of only
0.010, against a weighted national average of 0.097 ($n=244$).
For credit card and composite digital payment outcomes, the
most excluded profile converges on older, non-employed,
less-educated respondents, with predicted adoption
probabilities of 0.315 and 0.316 against national averages of
0.595 ($n=671$) and 0.608 ($n=693$) respectively.

Across technologies, digital payment exclusion operates through
overlapping socioeconomic disadvantages (especially age,
education, and weak economic attachment) rather than any single
barrier, while the most excluded subpopulations differ
meaningfully across payment instruments.

\section{Q.4: Digital literacy and mechanisms}
\label{sec:IRT}

This section uses the digital literacy score from
Section~\ref{sec:irt_method} to assess whether gaps in digital adoption
reflect broad differences in digital competence or more specific
domain-level deficits. The score is constructed from 20 binary CIUS
items spanning information-seeking, software and file management, and
security and privacy tasks.

The diagnostics support the use of the general-factor score as our main
summary measure: the ratio of the first to second tetrachoric
eigenvalue is 5.18 and McDonald's $\omega_h=0.76$, indicating a
dominant general factor. The loading pattern is also substantively
coherent: software and file-management items load most strongly on the
general factor, while security and privacy items retain an additional
domain-specific component. A scree plot, full loading estimates, and
additional dimensionality diagnostics are reported in
Appendix~\ref{app:irt}.

\subsection{Distributional patterns in digital literacy}

\begin{figure}[h!]
	\centering
	\caption{Digital Literacy Domain Scores by Age Group}
	\includegraphics[width=1\linewidth, height=0.5\textheight,
	keepaspectratio]{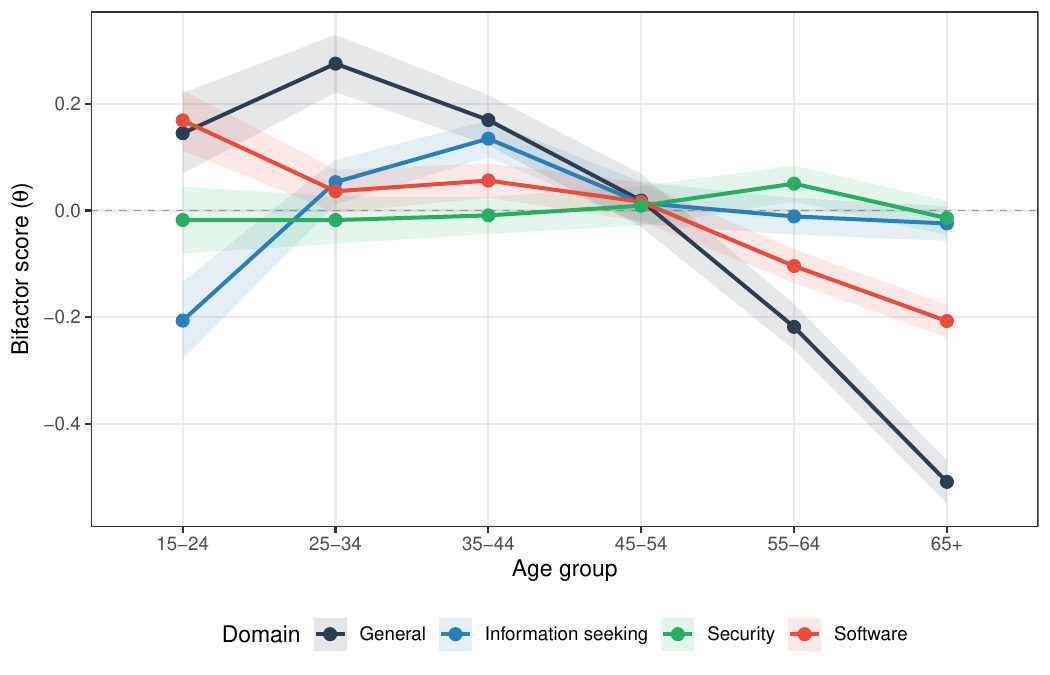}
	\label{fig:irt_age}
\footnotesize{\textit{Notes}: Survey-weighted bifactor domain scores by
		age group. Points are weighted means and bands denote 95\%
		confidence intervals. The dashed line at zero denotes the
		population average.}
\end{figure}

Figure~\ref{fig:irt_age} plots survey-weighted domain scores by age
group. The general score exhibits a pronounced age gradient: the
25--34 group records the highest mean (0.275), while the 65-plus group
records the lowest ($-0.509$), a gap of 0.784 standard deviations. The
decline is modest through middle age but steepens markedly after age
55, driven largely by the software and file-management dimension, where
scores fall below zero for the 55--64 group ($-0.104$) and decline
further for the 65-plus group ($-0.207$).

The information-seeking dimension tells a different story. The youngest
cohort (15--24) records the \emph{lowest} information-seeking score
($-0.206$), while the 35--44 group records the highest (0.135). This
suggests that younger Canadians may be highly engaged digitally without
being equally proficient in structured information-search tasks ---
locating services, comparing options, or navigating institutional
websites --- at which middle-aged respondents perform better.

\begin{figure}[h!]
	\caption{Security Subscore by Demographic Group}
	\includegraphics[width=1.2\linewidth, height=0.75\textheight,
	keepaspectratio]{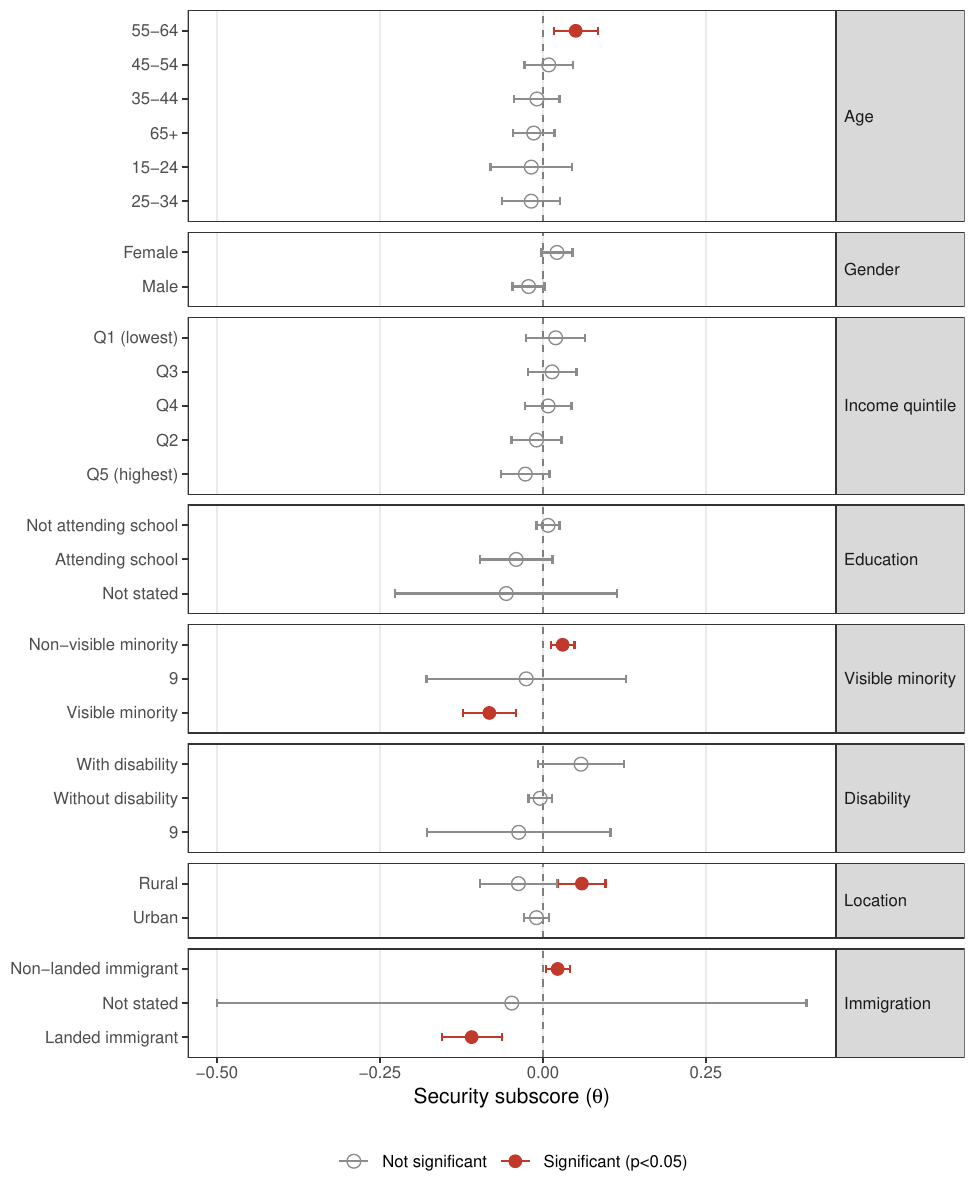}
	\label{fig:irt_security}
	
	\footnotesize{\textit{Notes}: Survey-weighted bifactor security
		subscore. Filled circles denote estimates significant at
		$p<0.05$; open circles denote non-significant estimates.}
\end{figure}

Security behavior exhibits a weaker age gradient. Most age groups
cluster near the population average, but the 55--64 cohort scores
significantly above zero on the security subscore
(0.050, 95\%~CI [0.016, 0.084]), possibly reflecting workplace exposure
to formal IT security practices during the organizational expansion of
the late 1990s and early 2000s. Figure~\ref{fig:irt_security} shows
that the security dimension does not follow a simple socioeconomic
gradient: visible minorities and landed immigrants score significantly
lower, whereas rural residents score significantly higher. These
patterns suggest that digital-inclusion policies should address
privacy-risk awareness and protective practices, not only access and
general skills.

Digital literacy itself is also unequally distributed across
income. Applying the income-ranked concentration index of
Section~\ref{sec:ci_method} to the rescaled literacy score
$\hat{L}_i$ (Panel~A of Table~\ref{tab:literacy_ci}) yields
$\widehat{\mathcal{C}}_L = 0.029$ ($\mathrm{SE} = 0.003$,
$p < 0.001$), confirming a statistically significant pro-rich
gradient in digital competence and motivating the
literacy-ranked analysis below.

\subsection{Digital literacy, the education gradient, and literacy-ranked concentration}
\label{sec:literacy_conditioning}

The persistent education gradient in the sequential logit may reflect
differences in digital capability, connectivity quality, or behavioral
preferences. To assess the capability channel, we augment the
sequential logit with the general digital literacy score and compare
the AME of the high-school-or-less indicator before and after
conditioning.

Table~\ref{tab:literacy_conditioning} reports results for Stages~1
and~3. At Stage~1 (internet use), the baseline education gap is about
5.5 percentage points; after conditioning on literacy, the effect
becomes statistically indistinguishable from zero, indicating that the
initial education gradient largely reflects differences in digital
capability. At Stage~3 (online banking), conditioning on literacy
reduces the gap by 61\%, but the effect remains significant at the
5\% level, pointing to behavioral and institutional mechanisms ---
trust, financial interface familiarity, and perceived risk --- that
continue to shape adoption beyond basic skill. The literacy score
itself raises the probability of online banking by 7.2 percentage
points per standard deviation.

These are descriptive conditioning effects rather than causal
mediation estimates, since digital literacy is jointly determined with
education and other socioeconomic characteristics. Even so, the
attenuation pattern is informative: capability accounts for most of
the education gradient at entry, while behavioral and institutional
frictions dominate at higher rungs.

Table~\ref{tab:literacy_ci} reports the literacy-ranked concentration
indices defined in Section~\ref{sec:ci_method} for six digital
activities. All six indices are positive, indicating that digital
participation is systematically tilted toward the more digitally
literate. Near-universal activities show the smallest gradients:
email use ($\widehat{\mathcal{C}}_y^L = 0.026$) and smartphone use
($0.030$) are only mildly concentrated at the upper end of the
literacy distribution. By contrast, routine transactional activities
display moderate and tightly clustered gradients: online banking
($0.066$), credit-card use ($0.062$), and government online services
($0.067$) have very similar concentration indices, suggesting a common
competence threshold for mainstream digital transactions.

Virtual-wallet adoption stands apart. Its literacy-ranked concentration
index is $\widehat{\mathcal{C}}_y^L = 0.344$, with a Murphy--Topel
standard error of $0.023$ and a 95\% confidence interval of
$[0.298,\;0.389]$, against a base rate of only 12.3\%. Part of this
larger value reflects the $1/\bar{y}$ scaling of the concentration
index for relatively rare outcomes, but the contrast with credit-card
use --- a similarly payment-oriented activity with a much higher base
rate and an index of only $0.062$ --- still points to a substantially
stronger literacy gradient for newer payment instruments. Taken
together, these results place digital competence as a distributional
link between economic resources and digital financial inclusion, complementing the
income-ranked indices in Section~\ref{sec:CI} and the conditioning
evidence above.

\begin{table}[htbp]
	\begin{center}
	\caption{Education gradient before and after conditioning on digital literacy}
	\label{tab:literacy_conditioning}
		\small
		\begin{tabular}{lccc}
			\toprule
			Stage & Model & AME ({HS or less}) & Reduction \\
			\midrule
			Internet use (Stage 1) & Baseline   & $-0.0545^{***}$ & --- \\
			Internet use (Stage 1) & + Literacy & $0.0000$        & 100\% \\
			\addlinespace
			Online banking (Stage 3) & Baseline   & $-0.0548^{***}$ & --- \\
			Online banking (Stage 3) & + Literacy & $-0.0213^{**}$  & 61\% \\
			\bottomrule
		\end{tabular}
		\end{center}
			\footnotesize
			\textit{Notes}: Entries report average marginal effects from
			survey-weighted sequential logit models. The literacy score is
			standardized to mean zero and unit variance. Reduction is
			the percentage change in the AME of {HS or less} after
			conditioning on literacy.
			$^{***}p<0.01$, $^{**}p<0.05$.
		
\end{table}

\begin{table}[H]
	\small
	\begin{center}
		\caption{Digital Literacy and Literacy-Ranked Concentration Indices}
		\label{tab:literacy_ci}
		\begin{tabular}{lcccc}
			\toprule
			\multicolumn{5}{l}{\textbf{Panel A. Income-Ranked Concentration of Digital Literacy}} \\
			\midrule
			Outcome & Mean & $\widehat{\mathcal{C}}_L$  & SE & 95\% CI \\
			\midrule
			Digital literacy score & 0.601 & 0.029 & 0.003 & $[0.023,\;0.034]$ \\
			\addlinespace
			\midrule
			\multicolumn{5}{l}{\textbf{Panel B. Literacy-Ranked Concentration of Digital Activities}} \\
			\midrule
			Outcome & Mean & $\widehat{\mathcal{C}}_y^L$ & SE & 95\% CI \\
			\midrule
			Email use                  & 0.957 & 0.026 & 0.002 & $[0.022,\;0.030]$ \\
			Online banking             & 0.862 & 0.066 & 0.004 & $[0.059,\;0.074]$ \\
			Virtual wallet             & 0.123 & 0.344 & 0.023 & $[0.298,\;0.389]$ \\
			Credit card                & 0.793 & 0.062 & 0.005 & $[0.053,\;0.071]$ \\
			Government online services & 0.851 & 0.067 & 0.004 & $[0.059,\;0.074]$ \\
			Smartphone use             & 0.945 & 0.030 & 0.002 & $[0.026,\;0.033]$ \\
			\bottomrule
		\end{tabular}
	\end{center}
	
	\footnotesize
\textit{Notes}: Panel A reports the income-ranked concentration
index $\widehat{\mathcal{C}}_L$ applied to the rescaled digital
literacy score $\hat{L}_i$, measuring the extent to which
digital competence is concentrated among higher-income
respondents; its standard error uses the closed-form
linearization variance estimator \citep{Kakwani1997},
treating the estimated literacy scores as fixed. Panel B
reports literacy-ranked concentration indices for selected
digital activities, where
$\widehat{\mathcal{C}}_y^L =
\frac{2}{\bar{y}\,W}
\sum_i w_i (y_i-\bar{y})(\hat{r}_i^L-\tfrac{1}{2})$
and $\hat{r}_i^L$ is the weighted midpoint fractional rank in
the distribution of $\hat{L}_i$. Standard errors in Panel B
account for both second-stage sampling variation and first-stage
IRT estimation uncertainty via a Murphy--Topel-type correction \citep{MurphyTopel1985}; see
Appendix~\ref{app:variance} for details. All estimates use
person weights \texttt{WTPG}.
\end{table}
\section{Conclusion}
\label{sec:conc}

This paper studies digital inequality in Canada using four
complementary empirical approaches: survey-weighted logistic
Lasso for digital financial adoption, an exact Shapley
decomposition of age--education gaps, a sequential logit tracing
where disadvantage emerges along the adoption path, and a
bifactor IRT measure of digital literacy. The results show that
digital inequality is multidimensional: income, education,
disability, age, and digital capability matter in different ways
and at different stages of participation.

Three conclusions stand out. First, education is the only
determinant that remains significant throughout the full adoption
ladder; this effect is only partly accounted for by income,
health, disability, preferences, and measured digital literacy,
pointing to behavioral and institutional frictions that persist
beyond measurable competence. Second, income-based inequality is
most pronounced for virtual-wallet adoption, where both the
income and literacy gradients are especially steep, and for
online banking the concentration reflects a broad socioeconomic
gradient in which employment and education are nearly as
important as income itself. Third, the disability penalty is
stage-specific rather than uniform: it disappears at online
banking but reappears with the largest penalty at the
digital-payments stage, pointing to accessibility gaps in retail
payment interfaces that banking platforms have largely addressed.
The digital literacy analysis further reveals that general
capability and security behavior follow different social
patterns, with security deficits concentrated among recent
immigrants and visible minorities rather than among seniors ---
a finding with direct implications for digital-inclusion policy.

\paragraph*{Limitations.}
Three limitations warrant acknowledgment. The CIUS 2020 covers
only the ten provinces, excluding the territories and First
Nations reserves, so the findings do not generalize to those
populations. The digital literacy score is estimated on the
routed subsample of internet users, which may introduce selection
into capability measurement and limits the comparability of
literacy-conditioned results with the full-population analyses.
The decomposition and literacy-conditioning exercises are
associational rather than causal; they identify plausible
mechanisms but cannot rule out confounding. Future work using
the CIUS 2022 wave could assess whether the stage-specific
barriers documented here have persisted or shifted in the
post-pandemic period.
	\newpage
	\bibliographystyle{chicago}
	\bibliography{JMT.bib}
	
\appendix

\section{Supplementary results for survey-weighted logistic Lasso (Q.1)}\label{app:svylasso}
\subsection{Debiased Lasso inference}
\label{app:inference}

From the weighted log-likelihood defined in \eqref{eq:logitlik}, the
score, information matrix, and negative Hessian are
\begin{align*}
	S(\theta)
	&=
	n^{-1}\sum_{i=1}^n
	w_i x_i\!\left(y_i-\Lambda(x_i'\theta)\right), \\
	I(\theta)
	&=
	n^{-1}\sum_{i=1}^n
	w_i^2 x_i x_i'
	\Lambda(x_i'\theta)\bigl(1-\Lambda(x_i'\theta)\bigr), \\
	H(\theta)
	&=
	n^{-1}\sum_{i=1}^n
	w_i x_i x_i'
	\Lambda(x_i'\theta)\bigl(1-\Lambda(x_i'\theta)\bigr).
\end{align*}

To conduct inference on individual components of $\theta$, we rely on
the following result: under regularity conditions and for any fixed unit
vector $\tau$, the debiased estimator $\tilde{\theta}^{\mathsf{DB}}$ in
\eqref{eq:DB} satisfies
\begin{equation*}
	n^{1/2}
	\left(
	\tau'
	H(\hat{\theta})^{-1}
	I(\hat{\theta})
	H(\hat{\theta})^{-1}
	\tau
	\right)^{-1/2}
	\tau'
	\bigl(\tilde{\theta}^{\mathsf{DB}}-\theta_0\bigr)
	\cond
	\mathcal{N}(0,1);
\end{equation*}
see \cite{JMT2026} for formal arguments.  
We also report debiased estimators of average marginal effects (AMEs).
For the binary regressor $x_{ij}$, the individual marginal effect is
defined as
\begin{equation*}
	\mathrm{ME}_{ij}(\theta)
	:=
	\Lambda(x_i'\theta)\big|_{x_{ij}=1}
	-
	\Lambda(x_i'\theta)\big|_{x_{ij}=0}.
\end{equation*}
The corresponding weighted sample AME is
\begin{equation*}
	\widehat{\mathrm{AME}}_j(\theta)
	:=
	\frac{1}{\sum_{i=1}^n w_i}
	\sum_{i=1}^n
	w_i\,\mathrm{ME}_{ij}(\theta).
\end{equation*}
The debiased estimator of the AME$_j$, the average marginal effect
of regressor $j$, is obtained by a one-step correction:
\begin{equation*}
	\widetilde{\mathrm{AME}}_j^{\mathsf{DB}}
	=
	\widehat{\mathrm{AME}}_j(\hat{\theta})
	+
	\nabla_{\theta}\widehat{\mathrm{AME}}_j(\hat{\theta})'
	H(\hat{\theta})^{-1}
	S(\hat{\theta}),
\end{equation*}
where
\begin{align*}
	\nabla_{\theta}\widehat{\mathrm{AME}}_j(\hat{\theta})
	&=
	\frac{1}{\sum_{i=1}^n w_i}
	\sum_{i=1}^n w_i
	\Bigl(
	x_i\,\Lambda(x_i'\hat{\theta})\bigl(1-\Lambda(x_i'\hat{\theta})\bigr)
	\big|_{x_{ij}=1}
	-
	x_i\,\Lambda(x_i'\hat{\theta})\bigl(1-\Lambda(x_i'\hat{\theta})\bigr)
	\big|_{x_{ij}=0}
	\Bigr).
\end{align*}
An asymptotic confidence interval for $\mathrm{AME}_j$ is based on
\begin{equation*}
	n^{1/2}
	\left(
	\nabla_{\theta}\widehat{\mathrm{AME}}_j(\hat{\theta})'
	H(\hat{\theta})^{-1}
	I(\hat{\theta})
	H(\hat{\theta})^{-1}
	\nabla_{\theta}\widehat{\mathrm{AME}}_j(\hat{\theta})
	\right)^{-1/2}
	\bigl(
	\widetilde{\mathrm{AME}}_j^{\mathsf{DB}}-\mathrm{AME}_j
	\bigr)
	\cond
	\mathcal{N}(0,1).
\end{equation*}
\subsection{Additional regression tables}
\label{app:extra_tables}

This appendix contains the full \texttt{svy LLasso} results for
internet use (Table~\ref{tab:lasso_internet_pumf}), email use
(Table~\ref{tab:lasso_email_pumf}), and separate virtual wallet
and credit card results (Tables~\ref{tab:lasso_vw_pumf} and
\ref{tab:lasso_cc_pumf}).
\renewcommand{\arraystretch}{0.75}
\begin{table}[htbp]
	\small
	\begin{threeparttable}
		\caption{Lasso Logistic Regression Results: Internet Use}
		\label{tab:lasso_internet_pumf}
		\begin{tabular}{llLLrLr}
			\toprule
			\multicolumn{1}{l}{Variables} &
			\multicolumn{1}{l}{Categories} &
			\multicolumn{1}{l}{$\texttt{svy LLasso}$} &
			\multicolumn{1}{l}{$\tilde{\theta}^{\mathsf{DB}}$} &
			\multicolumn{1}{l}{p-value} &
			\multicolumn{1}{l}{$\widetilde{\mathrm{AME}}^{\mathsf{DB}}$} &
			\multicolumn{1}{l}{p-value} \\
			\midrule
			\textit{Intercept} &  & 3.77 & 3.33^{***} & $<0.001$ & \multicolumn{1}{c}{-} & - \\
			\textit{Location} & Rural & -0.23 & -0.36^{***} & $<0.001$ & -0.02^{***} & $<0.001$ \\
			\textit{Age} & 15--24 & 0.37 & 1.01^{***} & $<0.001$ & 0.05^{***} & 0.001 \\
			& 25--34 & \multicolumn{1}{c}{-} & 0.57^{**} & 0.018 & 0.03^{**} & 0.043 \\
			& 35--44 & \multicolumn{1}{c}{-} & 0.49^{**} & 0.028 & 0.03^{*} & 0.057 \\
			& 55--64 & -0.64 & -0.51^{***} & 0.004 & -0.03^{**} & 0.017 \\
			& 65 and older & -1.61 & -1.30^{***} & $<0.001$ & -0.09^{***} & $<0.001$ \\
			\textit{Gender} & Female & \multicolumn{1}{c}{-} & 0.12 & 0.141 & 0.01 & 0.153 \\
			\textit{Aboriginal identity} & Aboriginal & \multicolumn{1}{c}{-} & -0.41^{*} & 0.055 & -0.03^{**} & 0.038 \\
			\textit{Language} & English & \multicolumn{1}{c}{-} & 0.24 & 0.612 & 0.01 & 0.615 \\
			& French & -0.49 & -0.04 & 0.936 & 0.00 & 0.951 \\
			& Eng and Fr & 0.07 & 0.60 & 0.220 & 0.03 & 0.269 \\
			\textit{Employment} & Employed & 0.44 & 0.49^{***} & $<0.001$ & 0.03^{***} & $<0.001$ \\
			\textit{Education} & HS or less & -0.89 & -0.93^{***} & $<0.001$ & -0.06^{***} & $<0.001$ \\
			& University degree & 0.32 & 0.42^{***} & $<0.001$ & 0.02^{***} & $<0.001$ \\
			\textit{Minority} & Visible minority & -0.03 & -0.47^{***} & 0.003 & -0.03^{***} & 0.002 \\
			\textit{Household type} & Family w/o child & \multicolumn{1}{c}{-} & -0.03 & 0.860 & 0.00 & 0.865 \\
			& Single & -0.54 & -0.64^{***} & $<0.001$ & -0.04^{***} & 0.001 \\
			& Other household & \multicolumn{1}{c}{-} & 0.30 & 0.345 & 0.02 & 0.397 \\
			\textit{Income} & Income Q1 & -0.55 & -0.45^{***} & $<0.001$ & -0.03^{***} & $<0.001$ \\
			& Income Q3 & \multicolumn{1}{c}{-} & 0.07 & 0.576 & 0.00 & 0.592 \\
			& Income Q4 & 0.01 & 0.27^{*} & 0.060 & 0.01^{*} & 0.081 \\
			& Income Q5 & 0.18 & 0.51^{***} & 0.003 & 0.03^{***} & 0.005 \\
			\textit{Immigration} & Non-landed & \multicolumn{1}{c}{-} & -0.05 & 0.797 & 0.00 & 0.805 \\
			\textit{Disability} & Disabled & -0.19 & -0.26^{**} & 0.037 & -0.02^{*} & 0.050 \\
			\textit{General health} & Excellent & \multicolumn{1}{c}{-} & 0.09 & 0.476 & 0.01 & 0.496 \\
			& Very good & 0.08 & 0.17^{*} & 0.083 & 0.01^{*} & 0.093 \\
			& Fair & \multicolumn{1}{c}{-} & -0.11 & 0.387 & -0.01 & 0.391 \\
			& Poor & -0.02 & -0.63^{***} & 0.003 & -0.04^{***} & $<0.001$ \\
			\textit{Province} & NL & \multicolumn{1}{c}{-} & -0.27 & 0.162 & -0.02 & 0.145 \\
			& PEI & \multicolumn{1}{c}{-} & -0.11 & 0.586 & -0.01 & 0.586 \\
			& NS & \multicolumn{1}{c}{-} & -0.32^{*} & 0.098 & -0.02^{*} & 0.081 \\
			& NB & \multicolumn{1}{c}{-} & -0.24 & 0.205 & -0.01 & 0.190 \\
			& QC & -0.32 & -0.73^{***} & $<0.001$ & -0.05^{***} & $<0.001$ \\
			& ON & \multicolumn{1}{c}{-} & -0.06 & 0.686 & 0.00 & 0.694 \\
			& MB & \multicolumn{1}{c}{-} & -0.48^{**} & 0.016 & -0.03^{***} & 0.008 \\
			& SK & \multicolumn{1}{c}{-} & -0.37^{*} & 0.058 & -0.02^{**} & 0.043 \\
			& BC & \multicolumn{1}{c}{-} & 0.09 & 0.623 & 0.00 & 0.640 \\
			\bottomrule
		\end{tabular}
		\begin{tablenotes}
			\footnotesize{\item \textit{Notes}: $n = 17,409$. ``-'' denotes variables not selected by \texttt{svy LLasso}. Reference: urban, age 45--54, male, non-Aboriginal, neither English nor French, not employed, some post-secondary, non-visible minority, family with child, income Q2, landed immigrant, not disabled, omitted health category, Alberta. Significance levels: *** $p<0.01$, ** $p<0.05$, * $p<0.10$.}
		\end{tablenotes}
	\end{threeparttable}
\end{table}

\renewcommand{\arraystretch}{0.75}
\begin{table}[htbp]
	\small
	\begin{threeparttable}
		\caption{Lasso Logistic Regression Results: Email Use}
		\label{tab:lasso_email_pumf}
		\begin{tabular}{llLLrLr}
			\toprule
			\multicolumn{1}{l}{Variables} &
			\multicolumn{1}{l}{Categories} &
			\multicolumn{1}{l}{$\texttt{svy LLasso}$} &
			\multicolumn{1}{l}{$\tilde{\theta}^{\mathsf{DB}}$} &
			\multicolumn{1}{l}{p-value} &
			\multicolumn{1}{l}{$\widetilde{\mathrm{AME}}^{\mathsf{DB}}$} &
			\multicolumn{1}{l}{p-value} \\
			\midrule
			\textit{Intercept} &  & 2.14 & -0.35 & 0.579 & \multicolumn{1}{c}{-} & - \\
			\textit{Location} & Rural & -0.16 & -0.32^{***} & 0.003 & -0.02^{***} & 0.003 \\
			\textit{Age} & 15--24 & \multicolumn{1}{c}{-} & 0.20 & 0.349 & 0.01 & 0.374 \\
			& 25--34 & 0.26 & 0.67^{***} & $<0.001$ & 0.03^{***} & 0.001 \\
			& 35--44 & 0.17 & 0.53^{***} & 0.003 & 0.03^{***} & 0.004 \\
			& 55--64 & -0.19 & -0.16 & 0.299 & -0.01 & 0.333 \\
			& 65 and older & -0.44 & -0.36^{**} & 0.026 & -0.02^{*} & 0.051 \\
			\textit{Gender} & Female & 0.10 & 0.22^{**} & 0.021 & 0.01^{**} & 0.022 \\
			\textit{Aboriginal identity} & Aboriginal & \multicolumn{1}{c}{-} & -0.34 & 0.205 & -0.02 & 0.156 \\
			\textit{Language} & English & 0.19 & 2.66^{***} & $<0.001$ & 0.26^{***} & $<0.001$ \\
			& French & \multicolumn{1}{c}{-} & 2.45^{***} & $<0.001$ & 0.09^{**} & 0.021 \\
			& Eng and Fr & 0.42 & 3.07^{***} & $<0.001$ & 0.15^{***} & $<0.001$ \\
			\textit{Employment} & Employed & 0.30 & 0.27^{**} & 0.025 & 0.02^{**} & 0.026 \\
			\textit{Education} & HS or less & -0.50 & -0.57^{***} & $<0.001$ & -0.04^{***} & $<0.001$ \\
			& University degree & 0.84 & 0.94^{***} & $<0.001$ & 0.05^{***} & $<0.001$ \\
			\textit{Minority} & Visible minority & -0.11 & -0.36^{**} & 0.030 & -0.02^{**} & 0.027 \\
			\textit{Household type} & Family w/o child & \multicolumn{1}{c}{-} & -0.10 & 0.502 & -0.01 & 0.505 \\
			& Single & -0.05 & -0.34^{**} & 0.026 & -0.02^{**} & 0.019 \\
			& Other household & \multicolumn{1}{c}{-} & -0.22 & 0.480 & -0.01 & 0.449 \\
			\textit{Income} & Income Q1 & -0.09 & -0.08 & 0.543 & -0.01 & 0.555 \\
			& Income Q3 & \multicolumn{1}{c}{-} & -0.06 & 0.666 & 0.00 & 0.664 \\
			& Income Q4 & \multicolumn{1}{c}{-} & -0.01 & 0.935 & 0.00 & 0.935 \\
			& Income Q5 & 0.27 & 0.35^{**} & 0.047 & 0.02^{**} & 0.036 \\
			\textit{Immigration} & Non-landed & 0.27 & 0.41^{**} & 0.022 & 0.03^{**} & 0.029 \\
			\textit{Disability} & Disabled & -0.15 & -0.47^{***} & 0.006 & -0.03^{***} & 0.004 \\
			\textit{General health} & Excellent & \multicolumn{1}{c}{-} & 0.10 & 0.497 & 0.01 & 0.509 \\
			& Very good & 0.04 & 0.13 & 0.270 & 0.01 & 0.276 \\
			& Fair & \multicolumn{1}{c}{-} & 0.05 & 0.758 & 0.00 & 0.763 \\
			& Poor & \multicolumn{1}{c}{-} & -0.03 & 0.926 & 0.00 & 0.926 \\
			\textit{Province} & NL & \multicolumn{1}{c}{-} & -0.10 & 0.635 & -0.01 & 0.625 \\
			& PEI & \multicolumn{1}{c}{-} & 0.09 & 0.677 & 0.01 & 0.689 \\
			& NS & \multicolumn{1}{c}{-} & -0.42^{*} & 0.051 & -0.03^{**} & 0.025 \\
			& NB & \multicolumn{1}{c}{-} & -0.56^{**} & 0.010 & -0.04^{***} & 0.002 \\
			& QC & -0.33 & -0.59^{***} & 0.009 & -0.04^{**} & 0.012 \\
			& ON & 0.07 & 0.10 & 0.556 & 0.01 & 0.556 \\
			& MB & \multicolumn{1}{c}{-} & -0.43^{*} & 0.057 & -0.03^{**} & 0.029 \\
			& SK & \multicolumn{1}{c}{-} & -0.27 & 0.219 & -0.02 & 0.180 \\
			& BC & 0.13 & 0.36^{*} & 0.074 & 0.02^{*} & 0.084 \\
			\bottomrule
		\end{tabular}
		\begin{tablenotes}
			\footnotesize{\item \textit{Notes}: $n = 15,153$. Reference: urban, age 45--54, male, non-Aboriginal, neither English nor French, not employed, some post-secondary, non-visible minority, family with child, income Q2, landed immigrant, not disabled, omitted health category, Alberta. Significance levels: *** $p<0.01$, ** $p<0.05$, * $p<0.10$.}
		\end{tablenotes}
	\end{threeparttable}
\end{table}

\renewcommand{\arraystretch}{0.75}
\begin{table}[htbp]
	\small
	\begin{threeparttable}
		\caption{Lasso Logistic Regression Results: Virtual Wallet}
		\label{tab:lasso_vw_pumf}
		\begin{tabular}{llLLrLr}
			\toprule
			\multicolumn{1}{l}{Variables} &
			\multicolumn{1}{l}{Categories} &
			\multicolumn{1}{l}{$\texttt{svy LLasso}$} &
			\multicolumn{1}{l}{$\tilde{\theta}^{\mathsf{DB}}$} &
			\multicolumn{1}{l}{p-value} &
			\multicolumn{1}{l}{$\widetilde{\mathrm{AME}}^{\mathsf{DB}}$} &
			\multicolumn{1}{l}{p-value} \\
			\midrule
			\textit{Intercept} &  & -2.05 & -3.62^{***} & $<0.001$ & \multicolumn{1}{c}{-} & - \\
			\textit{Location} & Rural & -0.16 & -0.56^{***} & $<0.001$ & -0.05^{***} & $<0.001$ \\
			\textit{Age} & 15--24 & 0.30 & 0.83^{***} & $<0.001$ & 0.11^{***} & $<0.001$ \\
			& 25--34 & 0.22 & 0.63^{***} & $<0.001$ & 0.08^{***} & $<0.001$ \\
			& 35--44 & \multicolumn{1}{c}{-} & 0.36^{***} & 0.003 & 0.04^{***} & 0.001 \\
			& 55--64 & -0.31 & -0.59^{***} & $<0.001$ & -0.06^{***} & $<0.001$ \\
			& 65 and older & -0.54 & -0.96^{***} & $<0.001$ & -0.08^{***} & $<0.001$ \\
			\textit{Gender} & Female & \multicolumn{1}{c}{-} & -0.05 & 0.521 & -0.01 & 0.516 \\
			\textit{Aboriginal identity} & Aboriginal & \multicolumn{1}{c}{-} & 0.08 & 0.740 & 0.01 & 0.731 \\
			\textit{Language} & English & \multicolumn{1}{c}{-} & 0.76 & 0.447 & 0.08 & 0.472 \\
			& French & \multicolumn{1}{c}{-} & 0.32 & 0.750 & 0.04 & 0.723 \\
			& Eng and Fr & \multicolumn{1}{c}{-} & 0.88 & 0.382 & 0.11 & 0.310 \\
			\textit{Employment} & Employed & \multicolumn{1}{c}{-} & 0.03 & 0.786 & 0.00 & 0.784 \\
			\textit{Education} & HS or less & \multicolumn{1}{c}{-} & -0.04 & 0.717 & 0.00 & 0.715 \\
			& University degree & 0.03 & 0.21^{**} & 0.030 & 0.02^{**} & 0.027 \\
			\textit{Minority} & Visible minority & 0.18 & 0.37^{***} & 0.003 & 0.04^{***} & 0.003 \\
			\textit{Household type} & Family w/o child & \multicolumn{1}{c}{-} & 0.08 & 0.463 & 0.01 & 0.456 \\
			& Single & \multicolumn{1}{c}{-} & 0.07 & 0.566 & 0.01 & 0.554 \\
			& Other household & \multicolumn{1}{c}{-} & 0.22 & 0.388 & 0.02 & 0.353 \\
			\textit{Income} & Income Q1 & \multicolumn{1}{c}{-} & 0.19 & 0.211 & 0.02 & 0.187 \\
			& Income Q3 & \multicolumn{1}{c}{-} & 0.21 & 0.130 & 0.02 & 0.111 \\
			& Income Q4 & \multicolumn{1}{c}{-} & 0.23^{*} & 0.099 & 0.03^{*} & 0.082 \\
			& Income Q5 & 0.25 & 0.68^{***} & $<0.001$ & 0.08^{***} & $<0.001$ \\
			\textit{Immigration} & Non-landed & \multicolumn{1}{c}{-} & 0.26^{*} & 0.059 & 0.03^{*} & 0.070 \\
			\textit{Disability} & Disabled & \multicolumn{1}{c}{-} & -0.01 & 0.945 & 0.00 & 0.944 \\
			\textit{General health} & Excellent & \multicolumn{1}{c}{-} & 0.22^{*} & 0.065 & 0.03^{*} & 0.053 \\
			& Very good & \multicolumn{1}{c}{-} & 0.05 & 0.624 & 0.01 & 0.618 \\
			& Fair & \multicolumn{1}{c}{-} & 0.00 & 0.993 & 0.00 & 0.993 \\
			& Poor & \multicolumn{1}{c}{-} & 0.39 & 0.240 & 0.05 & 0.182 \\
			\textit{Province} & NL & \multicolumn{1}{c}{-} & -0.24 & 0.234 & -0.02 & 0.265 \\
			& PEI & \multicolumn{1}{c}{-} & -0.11 & 0.621 & -0.01 & 0.629 \\
			& NS & \multicolumn{1}{c}{-} & -0.33 & 0.102 & -0.03 & 0.136 \\
			& NB & \multicolumn{1}{c}{-} & -0.26 & 0.222 & -0.03 & 0.255 \\
			& QC & \multicolumn{1}{c}{-} & -0.11 & 0.574 & -0.01 & 0.578 \\
			& ON & \multicolumn{1}{c}{-} & 0.01 & 0.940 & 0.00 & 0.939 \\
			& MB & \multicolumn{1}{c}{-} & -0.44^{**} & 0.025 & -0.04^{**} & 0.048 \\
			& SK & \multicolumn{1}{c}{-} & -0.18 & 0.370 & -0.02 & 0.391 \\
			& BC & \multicolumn{1}{c}{-} & 0.06 & 0.730 & 0.01 & 0.724 \\
			\bottomrule
		\end{tabular}
		\begin{tablenotes}
			\footnotesize{\item \textit{Notes}: $n = 12,124$. Reference: urban, age 45--54, male, non-Aboriginal, neither English nor French, not employed, some post-secondary, non-visible minority, family with child, income Q2, landed immigrant, not disabled, omitted health category, Alberta. Significance levels: *** $p<0.01$, ** $p<0.05$, * $p<0.10$.}
		\end{tablenotes}
	\end{threeparttable}
\end{table}

\renewcommand{\arraystretch}{0.75}
\begin{table}[htbp]
	\small
	\begin{threeparttable}
		\caption{Lasso Logistic Regression Results: Credit Card}
		\label{tab:lasso_cc_pumf}
		\begin{tabular}{llLLrLr}
			\toprule
			\multicolumn{1}{l}{Variables} &
			\multicolumn{1}{l}{Categories} &
			\multicolumn{1}{l}{$\texttt{svy LLasso}$} &
			\multicolumn{1}{l}{$\tilde{\theta}^{\mathsf{DB}}$} &
			\multicolumn{1}{l}{p-value} &
			\multicolumn{1}{l}{$\widetilde{\mathrm{AME}}^{\mathsf{DB}}$} &
			\multicolumn{1}{l}{p-value} \\
			\midrule
			\textit{Intercept} &  & 1.41 & 0.53 & 0.485 & \multicolumn{1}{c}{-} & - \\
			\textit{Location} & Rural & \multicolumn{1}{c}{-} & 0.07 & 0.368 & 0.01 & 0.390 \\
			\textit{Age} & 15--24 & -0.41 & -0.54^{***} & $<0.001$ & -0.09^{***} & $<0.001$ \\
			& 25--34 & \multicolumn{1}{c}{-} & 0.04 & 0.738 & 0.01 & 0.747 \\
			& 35--44 & \multicolumn{1}{c}{-} & 0.12 & 0.240 & 0.02 & 0.264 \\
			& 55--64 & \multicolumn{1}{c}{-} & -0.03 & 0.808 & 0.00 & 0.813 \\
			& 65 and older & \multicolumn{1}{c}{-} & -0.09 & 0.445 & -0.01 & 0.453 \\
			\textit{Gender} & Female & \multicolumn{1}{c}{-} & 0.01 & 0.888 & 0.00 & 0.892 \\
			\textit{Aboriginal identity} & Aboriginal & \multicolumn{1}{c}{-} & 0.18 & 0.361 & 0.03 & 0.396 \\
			\textit{Language} & English & 0.05 & 0.58 & 0.436 & 0.09 & 0.432 \\
			& French & -0.00 & -0.02 & 0.979 & 0.00 & 0.979 \\
			& Eng and Fr & \multicolumn{1}{c}{-} & 0.49 & 0.513 & 0.07 & 0.553 \\
			\textit{Employment} & Employed & 0.05 & 0.12 & 0.155 & 0.02 & 0.168 \\
			\textit{Education} & HS or less & -0.42 & -0.43^{***} & $<0.001$ & -0.07^{***} & $<0.001$ \\
			& University degree & 0.39 & 0.49^{***} & $<0.001$ & 0.07^{***} & $<0.001$ \\
			\textit{Minority} & Visible minority & -0.02 & -0.19^{*} & 0.080 & -0.03^{*} & 0.085 \\
			\textit{Household type} & Family w/o child & 0.09 & 0.34^{***} & $<0.001$ & 0.05^{***} & $<0.001$ \\
			& Single & \multicolumn{1}{c}{-} & 0.33^{***} & 0.001 & 0.05^{***} & 0.004 \\
			& Other household & \multicolumn{1}{c}{-} & 0.16 & 0.443 & 0.02 & 0.474 \\
			\textit{Income} & Income Q1 & -0.08 & -0.21^{*} & 0.065 & -0.03^{*} & 0.072 \\
			& Income Q3 & \multicolumn{1}{c}{-} & 0.06 & 0.596 & 0.01 & 0.610 \\
			& Income Q4 & \multicolumn{1}{c}{-} & 0.19^{*} & 0.076 & 0.03^{*} & 0.095 \\
			& Income Q5 & \multicolumn{1}{c}{-} & 0.13 & 0.250 & 0.02 & 0.274 \\
			\textit{Immigration} & Non-landed & \multicolumn{1}{c}{-} & 0.16 & 0.195 & 0.02 & 0.198 \\
			\textit{Disability} & Disabled & \multicolumn{1}{c}{-} & -0.44^{***} & 0.002 & -0.07^{***} & 0.001 \\
			\textit{General health} & Excellent & \multicolumn{1}{c}{-} & -0.21^{**} & 0.029 & -0.03^{**} & 0.029 \\
			& Very good & 0.00 & 0.01 & 0.876 & 0.00 & 0.880 \\
			& Fair & \multicolumn{1}{c}{-} & -0.07 & 0.587 & -0.01 & 0.593 \\
			& Poor & \multicolumn{1}{c}{-} & 0.36 & 0.154 & 0.05 & 0.206 \\
			\textit{Province} & NL & \multicolumn{1}{c}{-} & -0.33^{**} & 0.045 & -0.05^{**} & 0.036 \\
			& PEI & \multicolumn{1}{c}{-} & -0.04 & 0.797 & -0.01 & 0.801 \\
			& NS & \multicolumn{1}{c}{-} & -0.09 & 0.592 & -0.01 & 0.597 \\
			& NB & \multicolumn{1}{c}{-} & -0.16 & 0.335 & -0.03 & 0.332 \\
			& QC & -0.42 & -0.38^{**} & 0.014 & -0.06^{**} & 0.026 \\
			& ON & 0.06 & 0.24^{**} & 0.041 & 0.04^{**} & 0.049 \\
			& MB & \multicolumn{1}{c}{-} & 0.03 & 0.876 & 0.00 & 0.881 \\
			& SK & \multicolumn{1}{c}{-} & 0.01 & 0.964 & 0.00 & 0.966 \\
			& BC & \multicolumn{1}{c}{-} & 0.23 & 0.107 & 0.03 & 0.134 \\
			\bottomrule
		\end{tabular}
		\begin{tablenotes}
			\footnotesize{\item \textit{Notes}: $n = 12,124$. Reference: urban, age 45--54, male, non-Aboriginal, neither English nor French, not employed, some post-secondary, non-visible minority, family with child, income Q2, landed immigrant, not disabled, omitted health category, Alberta. Significance levels: *** $p<0.01$, ** $p<0.05$, * $p<0.10$.}
		\end{tablenotes}
	\end{threeparttable}
\end{table}
\subsection{Robustness to alternative tuning rules}
\label{app:robustness}

In the main text, $\lambda$ is selected by standard 10-fold
cross-validation with the AUC criterion via \texttt{glmnet}. Because
unequal selection probabilities raise a potential non-i.i.d.\ concern,
we consider two alternative tuning rules.

First, following \citet{Iparragirre2023}, we implement a
design-aware stratified cross-validation procedure in a PUMF-feasible
form, using weighted direct cross-validation (\texttt{dCV}) with
rescaled person weights. The full survey-design inputs required for
an exact implementation are not available in the PUMF analytical file;
we therefore use the final person weight rescaled to mean one, as
recommended by the CIUS User Guide \citep{CIUSUserGuide}.

Second, we implement the bootstrap-after-cross-validation (BCV) rule
of \citet{Chetverikov2025}. Non-constant regressors are standardized
using weighted means and standard deviations, and the BCV penalty
uses 1{,}000 bootstrap replications.

Both alternatives yield very similar results to the baseline.
After removing zero-variance, duplicate, and linearly dependent
columns, each model contains 46 candidate first-order regressors.
Under \texttt{dCV}, selected nonzero covariates number 41 (internet
use), 29 (online banking), 42 (email), 19 (virtual wallet), and 33
(credit card). Under BCV the corresponding numbers are 13, 6, 5, 7,
and 6. Despite these differences in sparsity, the substantive
conclusions are unchanged: strong age and education gradients, a
positive employment effect at online banking, a positive
visible-minority effect for virtual wallets, and a negative disability
effect for credit cards all persist across all three tuning approaches,
and the debiased AMEs remain very similar in magnitude.

\subsection{IIA tests}
\label{app:IIA}

We assessed whether the ``Not stated'' response should be retained as
a separate alternative using weighted Hausman--McFadden-type
multinomial logit tests comparing the full \{Yes, No, Not stated\}
specification with the restricted \{Yes, No\} specification. For
internet use, the test is not applicable (no observed ``Not stated''
responses in the PUMF estimation sample). For online banking, email,
and credit card use, the tests yield negative Hausman--McFadden
statistics; virtual wallet use produces a small positive statistic
(2.95). In all cases the IIA assumption is not rejected, supporting
the exclusion of ``Not stated'' responses from the binary
\texttt{svy LLasso} analysis.

\section{Supplementary decomposition results (Q.2)}
\subsection{Connectivity quality as an additional mediator}
\label{subsec:connectivity_mediator}

To examine the infrastructure channel in the age--education gap, we
augment the Q.2 decomposition with three mediators derived from the
CIUS: a \emph{mobile-only} access indicator, a below-CRTC-benchmark
speed indicator (50~Mbps), and a no-home-internet indicator.

\begin{figure}[H]
	\caption{Connectivity Quality as a Dimension of the Digital Divide}
	\includegraphics[width=0.95\linewidth, height=0.55\textheight,
keepaspectratio]{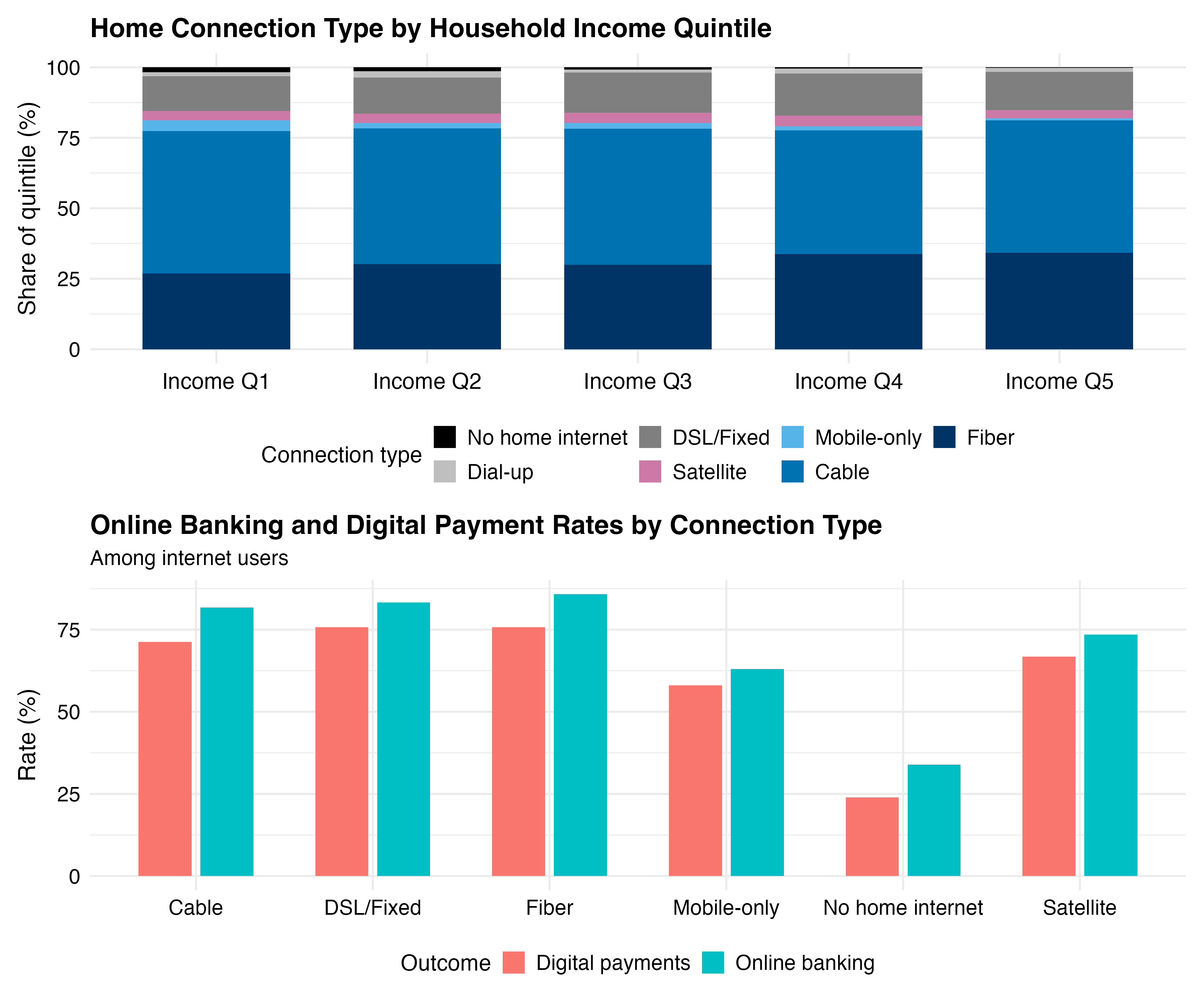}
	\label{fig:connectivity_quality}
	
	\footnotesize{\textit{Notes}: The upper panel shows the distribution of
		home connection types by household income quintile; the lower panel
		shows online banking and digital payment rates by connection type
		among internet users.}
\end{figure}

Figure~\ref{fig:connectivity_quality} shows clear differences in
digital engagement across connection types: online banking is highest
for fiber and cable users and substantially lower for mobile-only and
no-home-connection respondents. Connection quality varies across the
income distribution, but the gradient is modest relative to the
underlying income and education differences, and the decomposition
shows that connectivity quality explains only a limited additional
share of the residual education gradient. Digital inequality reflects
layered barriers, with infrastructure quality mattering at the margin
but not accounting for most of the remaining gap.

\subsection{Robustness of the decomposition across subsamples}
\label{app:decomp_robustness}

Table~\ref{tab:decomp_adj_subsamples} reports the adjusted decomposition
results for the urban non-disabled and urban samples. The qualitative
patterns closely match those in the full sample: adjustment reduces
the observed gaps, but substantial residual differences remain. 

\begin{table}[p]
	\begin{center}
	\caption{Unconditional Use by Age and Education: Raw Rates and Unadjusted Gaps}
	\label{tab:decomp_unadj}
		\small
		\begin{tabular}{lccc|ccc}
			\toprule
			\multicolumn{7}{c}{{Panel A: Unconditional rates}} \\
			\midrule
			& \multicolumn{3}{c}{Online banking}
			& \multicolumn{3}{c}{Health-information search} \\
			\cmidrule(lr){2-4} \cmidrule(lr){5-7}
			Age group
			& HS or less & Some PSE & University
			& HS or less & Some PSE & University \\
			\midrule
			15--24 & 0.638 & 0.817 & 0.977 & 0.726 & 0.711 & 0.887 \\
			25--44 & 0.829 & 0.901 & 0.915 & 0.751 & 0.828 & 0.912 \\
			45--64 & 0.640 & 0.805 & 0.872 & 0.567 & 0.728 & 0.856 \\
			65-plus & 0.381 & 0.584 & 0.710 & 0.381 & 0.623 & 0.760 \\
			\bottomrule
		\end{tabular}
		
		\begin{tabular}{lccc|ccc}
			\toprule
			\multicolumn{7}{c}{{Panel B: Unadjusted differences relative to reference cell}} \\
			\midrule
			& \multicolumn{3}{c}{Online banking}
			& \multicolumn{3}{c}{Health-information search} \\
			\cmidrule(lr){2-4} \cmidrule(lr){5-7}
			Age group
			& HS or less & Some PSE & University
			& HS or less & Some PSE & University \\
			\midrule
			15--24 & 0.257 & 0.436 & 0.596 & 0.345 & 0.330 & 0.506 \\
			25--44 & 0.448 & 0.521 & 0.535 & 0.370 & 0.447 & 0.531 \\
			45--64 & 0.260 & 0.424 & 0.491 & 0.186 & 0.347 & 0.475 \\
			65-plus & 0.000 & 0.203 & 0.329 & 0.000 & 0.242 & 0.379 \\
			\bottomrule
		\end{tabular}
\end{center}		
			\footnotesize
			\emph{Notes}: Panel A reports weighted unconditional probabilities of using online banking and searching for health information on the internet by age and education group. Panel B reports unadjusted differences relative to the reference cell (individuals aged 65 and older with high school education or less), corresponding to equation~\eqref{eq:m1}. All estimates use survey weights.
\end{table}
\begin{table}[htbp]
	\centering
	\caption{Exact Shapley Decomposition of Age--Education Gaps Across Urban Subsamples}
	\label{tab:decomp_adj_subsamples}
	\begin{threeparttable}
		\small
		\setlength{\tabcolsep}{4pt}
		
		\begin{tabular}{lrrrrrrr}
			\toprule
			\multicolumn{8}{l}{{Panel A. Urban Non-Disabled Sample}} \\
			\midrule
			Cell & Obs. gap & Adj. gap & Income & Disability & Health & Preferences & Total expl. \\
			\midrule
			15--24 $\times$ HS or less   & 0.257 & 0.055 & 0.001 & 0.000 & 0.002 & 0.187 & 0.190 \\
			15--24 $\times$ Some PSE     & 0.436 & 0.191 & 0.014 & 0.000 & -0.003 & 0.231 & 0.244 \\
			15--24 $\times$ University   & 0.596 & 0.380 & 0.011 & 0.000 & 0.000  & 0.189 & 0.200 \\
			25--44 $\times$ HS or less   & 0.448 & 0.223 & 0.001 & 0.000 & 0.004  & 0.224 & 0.229 \\
			25--44 $\times$ Some PSE     & 0.521 & 0.278 & 0.003 & 0.000 & 0.000  & 0.237 & 0.243 \\
			25--44 $\times$ University   & 0.535 & 0.262 & 0.011 & 0.000 & 0.003  & 0.230 & 0.244 \\
			45--64 $\times$ HS or less   & 0.260 & 0.101 & 0.013 & 0.000 & -0.001 & 0.145 & 0.152 \\
			45--64 $\times$ Some PSE     & 0.424 & 0.176 & 0.013 & 0.000 & 0.008  & 0.218 & 0.241 \\
			45--64 $\times$ University   & 0.491 & 0.226 & 0.015 & 0.000 & -0.003 & 0.239 & 0.253 \\
			65-plus $\times$ Some PSE    & 0.203 & 0.047 & 0.016 & 0.000 & -0.001 & 0.143 & 0.159 \\
			65-plus $\times$ University  & 0.329 & 0.100 & 0.024 & 0.000 & -0.005 & 0.180 & 0.195 \\
			
			\addlinespace
			\midrule
			\multicolumn{8}{l}{{Panel B. Urban Sample}} \\
			\midrule
			Cell & Obs. gap & Adj. gap & Income & Disability & Health & Preferences & Total expl. \\
			\midrule
			15--24 $\times$ HS or less   & 0.257 & 0.056 & 0.001 & -0.001 & 0.002 & 0.185 & 0.187 \\
			15--24 $\times$ Some PSE     & 0.436 & 0.192 & 0.014 & 0.002  & -0.003 & 0.229 & 0.242 \\
			15--24 $\times$ University   & 0.596 & 0.381 & 0.011 & 0.001  & 0.000  & 0.187 & 0.199 \\
			25--44 $\times$ HS or less   & 0.448 & 0.224 & 0.001 & -0.002 & 0.004  & 0.223 & 0.226 \\
			25--44 $\times$ Some PSE     & 0.521 & 0.279 & 0.003 & 0.001  & 0.000  & 0.236 & 0.240 \\
			25--44 $\times$ University   & 0.535 & 0.263 & 0.011 & 0.002  & 0.003  & 0.229 & 0.244 \\
			45--64 $\times$ HS or less   & 0.260 & 0.102 & 0.013 & -0.008 & -0.001 & 0.144 & 0.148 \\
			45--64 $\times$ Some PSE     & 0.424 & 0.177 & 0.013 & 0.007  & 0.008  & 0.216 & 0.244 \\
			45--64 $\times$ University   & 0.491 & 0.227 & 0.015 & 0.001  & -0.003 & 0.238 & 0.251 \\
			65-plus $\times$ Some PSE    & 0.203 & 0.048 & 0.016 & 0.005  & -0.001 & 0.142 & 0.160 \\
			65-plus $\times$ University  & 0.329 & 0.101 & 0.024 & -0.004 & -0.005 & 0.179 & 0.194 \\
			\bottomrule
		\end{tabular}
		
		\begin{tablenotes}[flushleft]
			\footnotesize
			\item \emph{Notes}: Reference cell is individuals aged 65 and older with high school or less education. Entries report exact simulation-based Shapley contributions from income, disability, health, and preference-related non-use. In Panel A, disability is excluded by sample restriction, so the disability contribution is zero by construction.
		\end{tablenotes}
		
	\end{threeparttable}
\end{table}
\clearpage
\section{Supplementary sequential logit results (Q.3)}
\label{app:seq_logit}

\subsection{Goodness-of-fit statistics}
\label{app:seq_gof}
Table~\ref{tab:gof_sequential} reports the full goodness-of-fit
statistics for the four sequential logit stages.
\begin{table}[!htbp]
	\small 
	\begin{center}
	\caption{Sequential Logit: Goodness-of-Fit Statistics}
	\label{tab:gof_sequential}
	
	\begin{tabular}{lrrrrrr}
		\toprule
		Stage & $n$ & McFadden $R^2$ & Dispersion & Resid.\ df & Null dev. & Resid.\ dev. \\
		\midrule
		S1: Internet use      & 17{,}409 & 0.29 & 1.08 & 17{,}362 & 9{,}495.9 & 6{,}713.9 \\
		S2: Email use         & 15{,}153 & 0.12 & 0.95 & 15{,}106 & 7{,}523.0 & 6{,}630.5 \\
		S3: Online banking    & 13{,}794 & 0.09 & 1.01 & 13{,}747 & 12{,}067.4 & 10{,}928.0 \\
		S4: Digital payments  & 10{,}559 & 0.06 & 0.99 & 10{,}512 & 9{,}241.9 & 8{,}676.8 \\
		\bottomrule
	\end{tabular}
\end{center}
	\footnotesize
	\textit{Notes}: The covariates correspond to the baseline specification used in Table~\ref{tab:lasso_bank_pumf}. Models are estimated using survey-weighted logit with person weights (\texttt{WTPG}). McFadden's pseudo-$R^2$ is defined as $1$-\text{Residual deviance}/\text{Null deviance}.
\end{table}
\subsection{Archetype and disadvantaged profiles}
\label{app:profiles}

Figure~\ref{fig:archetypes} reports the predicted conditional
probability of advancing at each rung for six representative
demographic archetypes constructed from the estimated sequential logit
models. The benchmark profile --- a middle-aged, employed,
English-speaking individual living in an urban family household with
children, holding some post-secondary education, and without a
disability --- exhibits high advancement probabilities at all rungs.
Profiles combining multiple sources of disadvantage show substantial
attrition, especially at the early access stage and at the transition
into online banking and digital payments.

\begin{table}[p]
	\centering
	
\begin{minipage}{0.95\linewidth}
	\centering
	\captionsetup{type=figure}
	\caption{Predicted Conditional Probabilities: Archetype Profiles}
	\includegraphics[width=\linewidth, height=0.52\textheight,
	keepaspectratio]{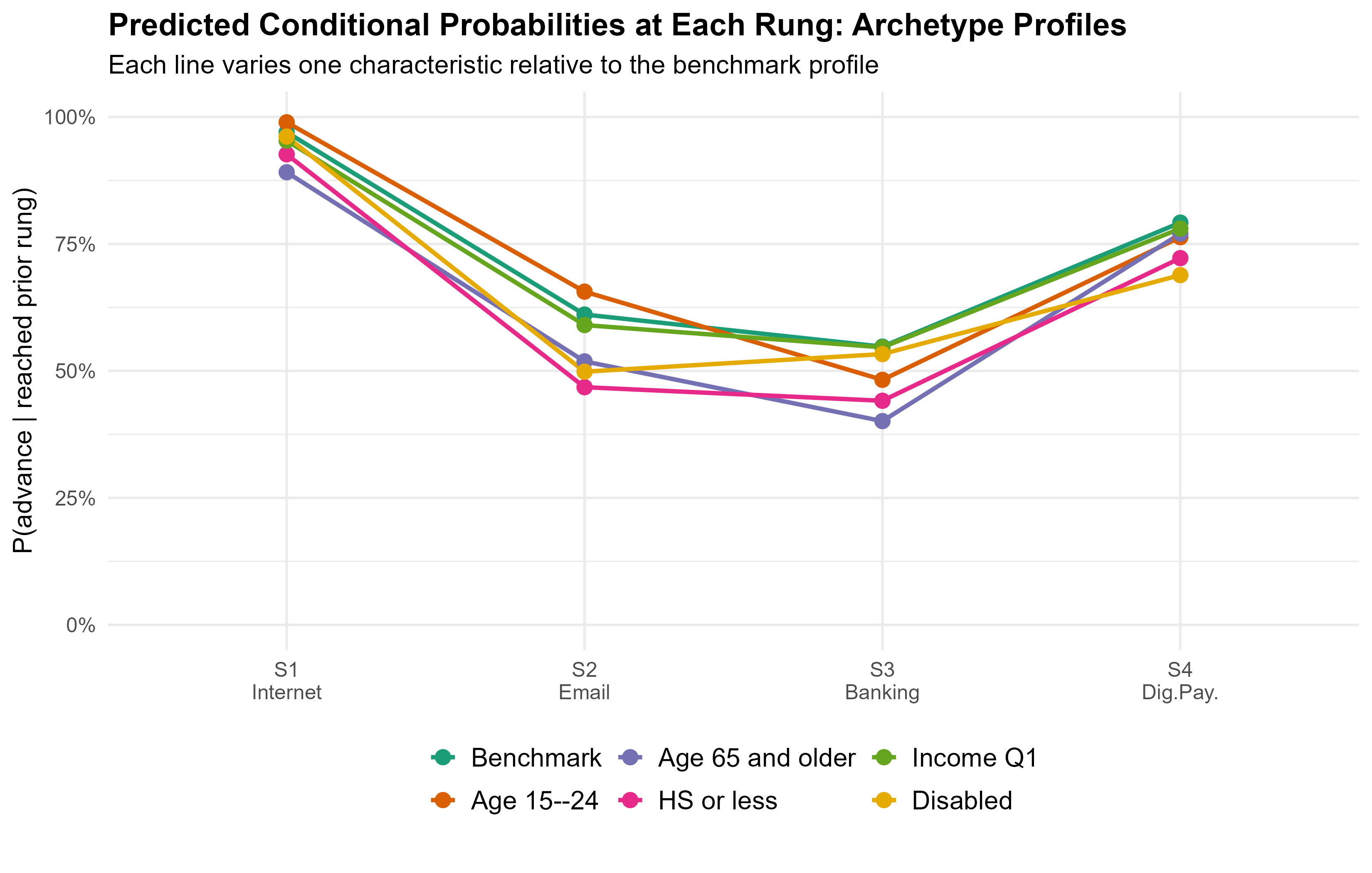}
	\label{fig:archetypes}
	\footnotesize
	\raggedright
	\textit{Notes}: Lines report predicted conditional probabilities
	$P(\text{advance}\mid\text{reached prior rung})$ from the four survey-weighted
	sequential logit models. Each archetype varies one characteristic
	relative to the benchmark profile; all other covariates are held
	at their reference categories.
\end{minipage}
	
	\vspace{1em}
	
	\begin{minipage}{0.95\linewidth}
		\centering
		\captionsetup{type=table}
		\caption{Characteristics of the Most Disadvantaged Profiles and Predicted Adoption Probabilities}
		\label{tab:disadvantaged_profiles_final}
		
		\begin{threeparttable}
			\small
			\begin{tabular}{l p{6.8cm} c c c}
				\toprule
				Outcome 
				& Key disadvantaged characteristics 
				& Prob. 
				& National average 
				& $n$ \\
				\midrule
				Virtual wallet &
				Age 65+, HS or less, income Q1, non-visible minority
				& 0.010 & 0.097 & 244 \\
				Credit card &
				Age 65+, not employed, HS or less, non-visible minority
				& 0.315 & 0.595 & 671 \\
				Digital payments &
				Age 65+, not employed, HS or less
				& 0.316 & 0.608 & 693 \\
				\bottomrule
			\end{tabular}
			
			\begin{tablenotes}[flushleft]
				\footnotesize
				\item \textit{Notes}: Key characteristics are the weighted modal characteristics
				among respondents in the bottom decile of predicted adoption probabilities,
				retaining those with modal share exceeding 0.80 (0.75 for virtual wallet).
				Predicted probabilities are evaluated at the corresponding profile-specific
				covariate vector using the survey-weighted sequential logit model; remaining
				covariates are set to their weighted sample modes. National averages are
				weighted sample-average predictions. For virtual wallet and credit card
				outcomes, non-visible minority status appears because visible minorities are
				over-represented in the upper tail of the virtual-wallet distribution (see
				Section~\ref{sec:digital_payments}). \emph{Digital payments} denotes adoption
				of either a virtual wallet or a credit card for online purchases.
			\end{tablenotes}
		\end{threeparttable}
	\end{minipage}
\end{table}
\newpage 
\section{IRT model details (Q.4)}
\label{app:irt}

\subsection{Item list}\label{app:irt_items}

The digital literacy score is constructed from 20 binary indicators
from the 2020 CIUS, grouped into three conceptual domains.

\medskip

\noindent\textbf{Information-seeking activities}
\begin{multicols}{2}
	\begin{itemize}[noitemsep, topsep=1pt, leftmargin=*]
		\item Using social networking services
		\item Making voice or video calls over the Internet
		\item Searching for community events
		\item Reading news or current affairs online
		\item Looking up locations or directions
		\item Searching for health information
		\item Researching goods or services before purchase
	\end{itemize}
\end{multicols}

\noindent\textbf{Software and file management}
\begin{multicols}{2}
	\begin{itemize}[noitemsep, topsep=1pt, leftmargin=*]
		\item Copying or moving files or folders
		\item Using word processing software
		\item Creating presentations
		\item Using spreadsheet software
		\item Editing photos or videos
		\item Deleting browser history
		\item Downloading files
		\item Uploading files to cloud storage
		\item Updating operating system software
	\end{itemize}
\end{multicols}

\noindent\textbf{Security and privacy}
\begin{multicols}{2}
	\begin{itemize}[noitemsep, topsep=0pt, leftmargin=*]
		\item Checking whether a website connection is secure
		\item Restricting location data sharing
		\item Refusing or limiting advertising tracking
		\item Changing privacy settings on online accounts
	\end{itemize}
\end{multicols}
\subsection{EAP estimator}\label{app:EAP}
The expected a posteriori (EAP) estimate (the general-factor score for individual $i$) is given by 
\begin{equation*}
\hat{\theta}_i^{(G)}
=
\frac{
	\int \theta^{(G)}\,
	p\left(\mathbf{y}_i \mid \theta^{(G)},
	\theta^{(D)}, \hat{\psi}\right)
	\phi\!\left(\theta^{(G)}\right)\,
	d\theta^{(G)}\,d\theta^{(D)}
}{
	\int
	p\left(\mathbf{y}_i \mid \theta^{(G)},
	\theta^{(D)}, \hat{\psi}\right)
	\phi\!\left(\theta^{(G)}\right)\,
	d\theta^{(G)}\,d\theta^{(D)}
},
\end{equation*}
where $\mathbf{y}_i = (y_{i1},\ldots,y_{i,20})'$, $\hat{\psi} = \{\hat{a}_j^{(G)}, \hat{a}_j^{(D_j)},
\hat{b}_j\}_{j=1}^{20}$, 
\begin{equation*}
p\left(\mathbf{y}_i \mid \theta^{(G)}, \theta^{(D)},
\hat{\psi}\right)
=
\prod_{j=1}^{20}
P\left(y_{ij}=1 \mid \theta^{(G)}, \theta_i^{(D_j)},
\hat{\psi}\right)^{y_{ij}}
\!\left[1-P\left(y_{ij}=1 \mid \theta^{(G)},
\theta_i^{(D_j)}, \hat{\psi}\right)\right]^{1-y_{ij}}
\end{equation*}
is the conditional likelihood of $\mathbf{y}_i$ given both latent
factors, $\phi(\cdot)$ is the standard normal prior, and $\theta^{(D)}$
collects the domain-specific factors integrated out jointly with
$\theta^{(G)}$. The integrals are evaluated numerically via EM or 
quasi-Monte Carlo EM algorithms.
\subsection{Bifactor loadings for the digital literacy score}\label{app:irt_loadings}
\begin{table}[h]
	\small
	\begin{center}
		\caption{Bifactor Loadings for the Digital Literacy Score}
		\label{tab:irt_loadings}
		\begin{tabular}{lcccc}
			\toprule
			Item & General & InfoSeek & Software & Security \\
			\midrule
			
			\multicolumn{5}{l}{\textit{Information-seeking}} \\
			
			Social networking       & 0.382 & 0.058 & 0.284 & 0.032 \\
			Voice/video calls       & 0.465 & 0.127 & 0.284 & 0.010 \\
			Community events        & 0.455 & -0.046 & 0.633 & -0.051 \\
			News                    & 0.483 & 0.062 & 0.405 & 0.016 \\
			Locations/directions    & 0.562 & 0.074 & 0.441 & 0.043 \\
			Health information      & 0.473 & -0.030 & 0.556 & 0.010 \\
			Goods/services research & 0.579 & 0.048 & 0.476 & 0.066 \\
			
			\addlinespace
			\multicolumn{5}{l}{\textit{Software and file management}} \\
			
			Copy/move files         & 0.765 & 0.440 & 0.002 & 0.082 \\
			Word processing         & 0.783 & 0.530 & -0.009 & -0.022 \\
			Presentations           & 0.687 & 0.465 & 0.020 & -0.043 \\
			Spreadsheet basics      & 0.717 & 0.503 & -0.027 & -0.028 \\
			Edit photo/video        & 0.635 & 0.315 & 0.069 & 0.083 \\
			Delete browser history  & 0.470 & 0.111 & 0.041 & 0.247 \\
			Download files          & 0.714 & 0.309 & 0.090 & 0.148 \\
			Upload to cloud         & 0.608 & 0.214 & 0.169 & 0.118 \\
			Update OS               & 0.569 & 0.132 & 0.097 & 0.262 \\
			
			\addlinespace
			\multicolumn{5}{l}{\textit{Security and privacy}} \\
			
			Check HTTPS             & 0.512 & 0.010 & 0.022 & 0.450 \\
			Restrict location data  & 0.589 & -0.005 & -0.008 & 0.572 \\
			Refuse ad tracking      & 0.610 & -0.021 & -0.023 & 0.627 \\
			Change privacy settings & 0.596 & 0.047 & 0.064 & 0.441 \\
			
			\bottomrule
		\end{tabular}
	\end{center}
	
	\footnotesize{
		\textit{Notes}: Schmid--Leiman standardized loadings from the weighted
		bifactor 2PL model estimated using normalized survey weights
		($n=12{,}065$).
		McDonald's $\omega_h=0.760$ and $\omega_{\text{total}}=0.782$.
	}
\end{table}
\subsection{Dimensionality diagnostics}\label{app:irt_dd}

\begin{figure}[h!]
	\centering
	\caption{Scree Plot of the Tetrachoric Correlation Matrix}
	\label{fig:scree_plot}
	
	\includegraphics[width=0.75\linewidth]{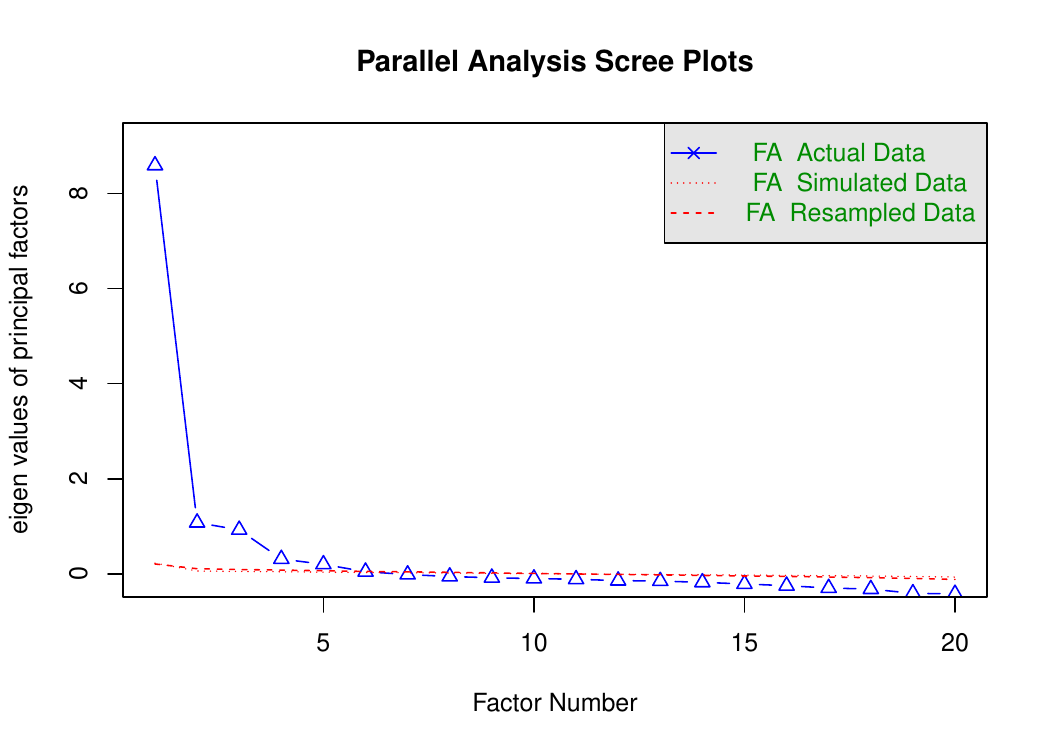}
	
	\vspace{0.5em}
	\begin{minipage}{0.75\linewidth}
		\footnotesize
		\textit{Notes}: Eigenvalues of the tetrachoric correlation matrix for the
		20 CIUS digital-skill items. The dashed reference lines correspond to
		parallel analysis based on factor analysis (FA), comparing eigenvalues
		from the observed data with those from simulated and resampled datasets.
		The ratio of the first to the second eigenvalue is 5.18, indicating a
		dominant general factor.
	\end{minipage}
\end{figure}

The ratio of the first to the second eigenvalue of the tetrachoric
correlation matrix is 5.18, indicating a strong dominant general
factor. McDonald's $\omega_h = 0.760$ indicates that approximately
three-quarters of the composite-score variance is attributable to the
general factor ($\omega_{\text{total}} = 0.782$). Figure~\ref{fig:scree_plot}
reports the scree plot.

\subsection{Cyber victimization and digital exclusion}
\label{sec:cyber_exclusion}

\begin{figure}[ht]
	\centering
	\caption{Cyber Victimization and Digital Financial Exclusion by Quintile of the Bifactor Security Score}
	\label{fig:cyber_inverse_concentration}
	
	\includegraphics[width=0.82\linewidth]{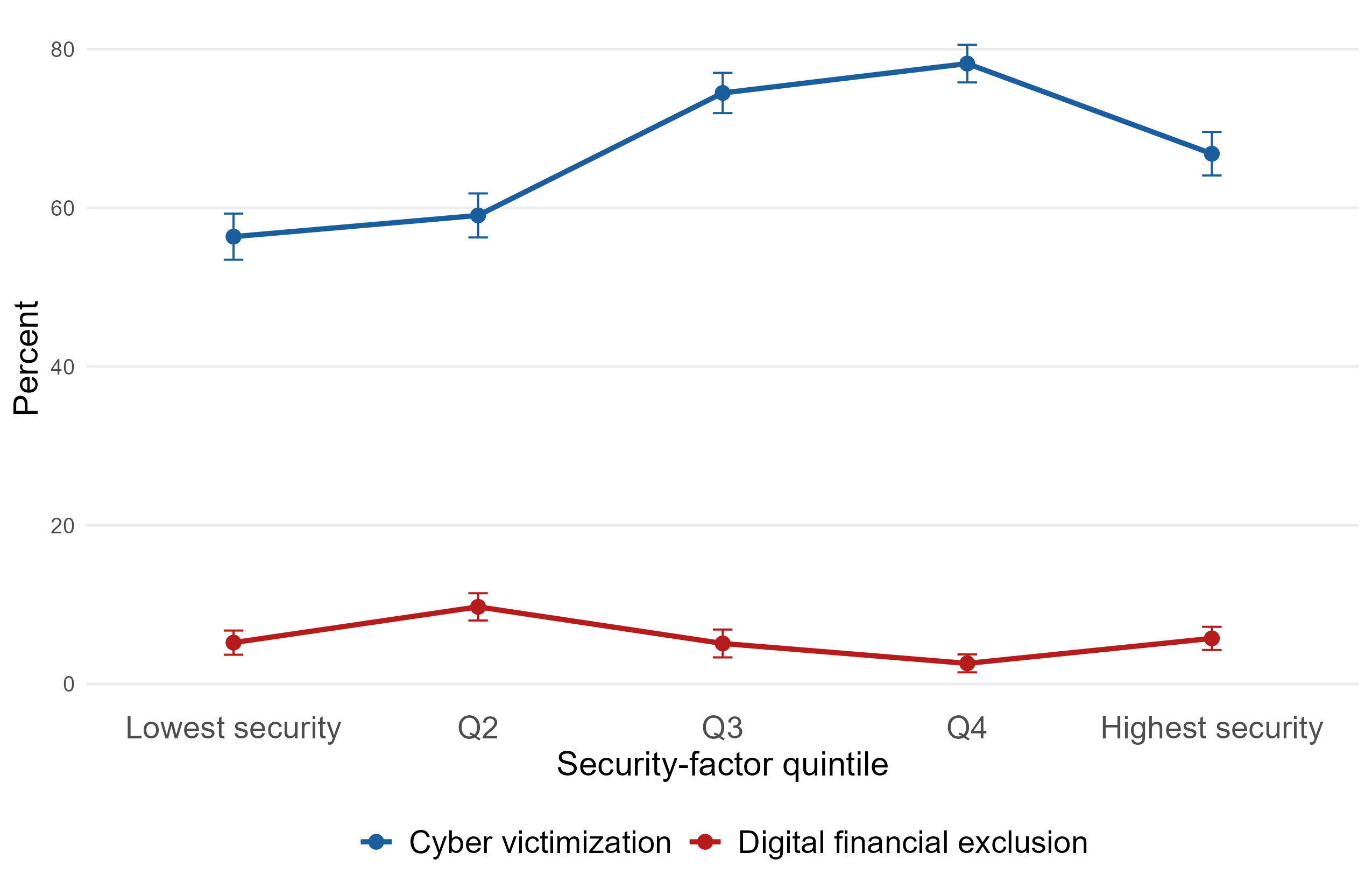}
	
	\vspace{0.5em}
	\begin{minipage}{0.82\linewidth}
		\footnotesize
		\textit{Notes}: Points show survey-weighted percentages; vertical bars indicate
		95\% confidence intervals. Cyber victimization is measured among internet
		users. Digital financial exclusion is defined as non-use of online banking
		and digital payment methods. Quintiles are from the standardized
		security factor derived from the bifactor IRT model.
	\end{minipage}
\end{figure}

Figure~\ref{fig:cyber_inverse_concentration} compares cyber
victimization and digital financial exclusion across quintiles of the
bifactor security score. Cyber victimization is higher in the middle
and upper part of the security distribution, whereas digital financial
exclusion remains low throughout and reaches its minimum in the
upper-security quintiles. This indicates that victimization is
concentrated among digitally active and exposed users rather than
among the most excluded. Higher general digital capability is
associated with both a greater probability of reporting a cyber
incident and a lower probability of digital financial exclusion; the
security-specific factor is positively associated with victimization
but has no meaningful association with exclusion. The inverse pattern
thus operates mainly through overall digital engagement rather than
security behavior alone.

\subsection{Variance estimation for literacy-ranked concentration indices}
\label{app:variance}

This appendix describes the variance estimator for the literacy-ranked
concentration indices in Table~\ref{tab:literacy_ci}. Because the
literacy rank $\hat r_i^L$ is constructed from an estimated bifactor
IRT model, $\widehat{\mathcal{C}}_y^L$ is a
two-step estimator and its sampling variance must account for
first-stage estimation uncertainty. We use a Murphy--Topel-style
correction \citep{MurphyTopel1985} to account for IRT parameter
uncertainty in the second-stage variance.

Let $\psi_0$ denote the true IRT parameter vector and $\hat{\psi}
=
\{\hat a_j^{(G)},\hat a_j^{(D_j)},\hat b_j\}_{j=1}^{20}$ be 
its first-stage estimate from the weighted bifactor 2PL model. Let $
\mathbf{y}_i=(y_{i1},\ldots,y_{i,20})'$
denote respondent $i$'s vector of binary IRT item responses. Let
$\hat\theta_i^{(G)}$ be respondent $i$'s EAP general-factor score and
$\hat r_i^L$ the corresponding weighted midpoint fractional rank,
satisfying $
\sum_{i=1}^n w_i \hat r_i^L/W = \tfrac{1}{2}$
by construction, where $W={\sum_{i=1}^n w_i}$. The survey-weighted concentration index
estimator is
\begin{equation}
	\label{eq:ci_def}
	\widehat{\mathcal C}_y^L
	=
	\frac{2}{\bar y\, W}
	\sum_{i=1}^nw_i
	(y_i-\bar y)
	\left(\hat r_i^L-\tfrac{1}{2}\right),
	\qquad
	\bar y=\sum_{i=1}^n w_i y_i/W.
\end{equation}
Let $\mu_y=\mathrm{E}[y_i]$ denote the population mean.
\paragraph*{Asymptotic linear representation.}

Let $U_i(\psi)$ denote the score contribution of observation $i$ from
the weighted bifactor 2PL likelihood, and let
\begin{equation*}
A
:=
-
\mathrm{E}\left[
\frac{\partial U_i(\psi_0)}{\partial \psi'}
\right]\ \text{and}\ B
:=
\frac{\partial \mathcal C_y^L}{\partial \psi'}
\end{equation*}
denote the expected information matrix, and the gradient of the population concentration index w.r.t. the first-stage parameters, respectively. Under standard regularity conditions for
two-step M-estimation, the first-stage estimator satisfies
\begin{equation*}
n^{1/2}(\hat\psi-\psi_0)
=
A^{-1}{n^{-1/2}}\sum_{i=1}^n U_i(\psi_0)+o_p(1),
\end{equation*}
and the asymptotic linear representation of the second-stage estimator
is
\begin{equation*}
n^{1/2}\bigl(\widehat{\mathcal C}_y^L-\mathcal C_y^L\bigr)
={n^{-1/2}}\sum_{i=1}^n \varphi_i^C
+
BA^{-1}{n^{-1/2}}\sum_{i=1}^n U_i(\psi_0)
+
o_p(1),
\end{equation*}
where $\varphi_i^C$ is the influence function of the concentration
index treating literacy ranks as known. Define
\begin{equation*}
V_C:=\mathrm{Var}(\varphi_i^C),\qquad
\Omega:=\mathrm{Var}(U_i),\qquad
\Gamma_{U\varphi}:=\mathrm{Cov}(U_i,\varphi_i^C).
\end{equation*}
Equivalently, the asymptotic covariance matrix of the first-stage
estimator is
\begin{equation*}
V_\psi=A^{-1}\Omega A^{-1}.
\end{equation*}
The Murphy--Topel-type asymptotic variance of
$n^{1/2}\,\widehat{\mathcal C}_y^L$ is therefore
\begin{equation}
	\label{eq:mt_full}
	V_{MT}
	=
	V_C
	+
	BA^{-1}\Omega A^{-1}B'
	+
	2BA^{-1}\Gamma_{U\varphi}
	=
	V_C
	+
	BV_\psi B'
	+
	2BA^{-1}\Gamma_{U\varphi}.
\end{equation}

\paragraph*{Influence function of the concentration index.}

Write the population concentration index as a smooth function of two
moments:
\begin{equation*}
	\mathcal C_y^L
	=
	g(\mu_{yr},\mu_y)
	:=
	\frac{2\mu_{yr}}{\mu_y}-1,
	\qquad
	\mu_{yr}=\mathrm{E}[y_i r_i^L],
	\qquad
	\mu_y=\mathrm{E}[y_i].
\end{equation*}
A first-order Taylor expansion of $g$ around $(\mu_{yr},\mu_y)$ yields
the influence function
\begin{equation*}
	\varphi_i^C
	=
	\frac{\partial g}{\partial \mu_{yr}}
	\bigl(y_i r_i^L-\mu_{yr}\bigr)
	+
	\frac{\partial g}{\partial \mu_y}
	\bigl(y_i-\mu_y\bigr),
\end{equation*}
up to a higher-order remainder term.
Since
\begin{equation*}
\frac{\partial g}{\partial \mu_{yr}}=\frac{2}{\mu_y},
\qquad
\frac{\partial g}{\partial \mu_y}
=
-\frac{2\mu_{yr}}{\mu_y^2},
\end{equation*}
it follows that
\begin{equation*}
\varphi_i^C
=
\frac{2}{\mu_y}\bigl(y_i r_i^L-\mu_{yr}\bigr)
-
\frac{2\mu_{yr}}{\mu_y^2}(y_i-\mu_y).
\end{equation*}
Using $
\mathcal C_y^L=\frac{2\mu_{yr}}{\mu_y}-1$, we get 
$\frac{\mu_{yr}}{\mu_y}=\frac{1+\mathcal C_y^L}{2},$
hence 
\begin{equation}
	\label{eq:ci_if}
	\varphi_i^C
	=
	\frac{y_i}{\mu_y}
	\left(2r_i^L-1-\mathcal C_y^L\right).
\end{equation}
One can verify directly that $\mathrm{E}[\varphi_i^C]=0$. The sample analogue, evaluated at estimated quantities, is
\begin{equation*}
\hat\varphi_i^C
=
\frac{w_i\, y_i}{W\,\bar{y}}
\left(2\hat r_i^L-1-\widehat{\mathcal C}_y^L\right).
\end{equation*}
The variance of
$n^{1/2}\,\widehat{\mathcal C}_y^L$ treating literacy ranks as known
is estimated by
\begin{equation*}
\widehat V_C
=
n\sum_{i=1}^n (\hat\varphi_i^C)^2.
\end{equation*}

\paragraph*{First-stage covariance and cross term.}
The estimated score matrix $
\widehat U
=
(\hat U_1',\ldots,\hat U_n')'$
is obtained from the fitted IRT model using
\texttt{estfun.AllModelClass()}. Let
$
\widehat\Omega
=
\frac{1}{n}\sum_{i=1}^n \hat U_i \hat U_i'$
denote the empirical covariance matrix of the first-stage score
contributions, and let $\widehat A^{-1}$ denote the estimated inverse
information matrix. Then the implied first-stage covariance matrix is
\begin{equation*}
\widehat V_\psi
=
\widehat A^{-1}\widehat\Omega \widehat A^{-1}.
\end{equation*}
In the implementation, we also extract \texttt{vcov()} from the fitted
IRT object as a numerical check on the first-stage covariance
calculation. The cross-covariance term in \eqref{eq:mt_full} is estimated by
\begin{equation}
	\label{eq:cov_u_phi}
	\widehat\Gamma_{U\varphi}
	=
	\frac{1}{n}\sum_{i=1}^n
	\hat U_i
	\left(\hat\varphi_i^C-\bar\varphi^C\right),
	\qquad
	\bar\varphi^C
	=
	\frac{1}{n}\sum_{i=1}^n \hat\varphi_i^C.
\end{equation}

\paragraph*{Kernel-smoothed Jacobian.}

The weighted midpoint rank $\hat r_i^L$ is a non-smooth step function
of the estimated literacy score, so $\partial \hat r_i^L/\partial\psi$
does not exist in the classical sense. To compute $\hat B$, we replace
the exact rank by the smooth approximation
\begin{equation}
	\label{eq:smoothed_rank}
	\tilde r_i^L(h)
	=
	\sum_{j=1}^n
	\frac{w_j}{W}
	\Phi\!\left(
	\frac{\hat\theta_i^{(G)}-\hat\theta_j^{(G)}}{h}
	\right),
\end{equation}
where $\Phi(\cdot)$ is the standard normal CDF and $h$ is a bandwidth
chosen by Silverman’s rule applied to
$\{\hat\theta_i^{(G)}\}_{i=1}^n$. The smoothed rank converges to the
exact midpoint rank as $h\to 0$ and is used only for differentiation;
the point estimate continues to use the exact rank. Its derivative
with respect to $\psi$ is
\begin{equation}
	\label{eq:smoothed_rank_deriv}
	\frac{\partial \tilde r_i^L(h)}{\partial \psi'}
	=
	\sum_{j=1}^n\frac{w_j}{Wh}\,
	\phi\!\left(
	\frac{\hat\theta_i^{(G)}-\hat\theta_j^{(G)}}{h}
	\right)
	\left(
	\frac{\partial \hat\theta_i^{(G)}}{\partial \psi'}
	-
	\frac{\partial \hat\theta_j^{(G)}}{\partial \psi'}
	\right),
\end{equation}
where $\phi(\cdot)$ is the standard normal density. The Jacobian of the EAP score satisfies
\begin{equation}
	\label{eq:eap_derivative}
	\frac{\partial \hat\theta_i^{(G)}}{\partial \psi}
	=
	\mathrm{Cov}\!\left(
	\eta^{(G)},\;
	\frac{\partial\log p(\mathbf{y}_i\mid \eta,\psi)}{\partial\psi}
	\;\middle|\;
	\mathbf{y}_i
	\right),
\end{equation}
which is evaluated numerically on a Monte Carlo quadrature grid. The
estimated gradient of the concentration index is then
\begin{equation}
	\label{eq:B_hat}
	\hat B
	=
	\frac{2}{\bar{y}\,W}
	\sum_{i=1}^n
	w_i
	(y_i-\bar y)
	\frac{\partial \tilde r_i^L(h)}{\partial \psi'}.
\end{equation}

\paragraph*{Final variance estimator.}

Substituting the sample analogues into \eqref{eq:mt_full}, the
implemented variance estimator is
\begin{equation}
	\label{eq:mt_hat}
	\widehat{\mathrm{Var}}
	\!\left(\widehat{\mathcal C}_y^L\right)
	=
	\frac{1}{n}
	\left(
	\widehat V_C
	+
	\hat B\,\widehat A^{-1}\widehat\Omega \widehat A^{-1}\hat B'
	+
	2\hat B\,\widehat A^{-1}\widehat\Gamma_{U\varphi}
	\right),
\end{equation}
and the corresponding standard error is
\begin{equation*}
{\mathrm{SE}}
=
\sqrt{
	\widehat{\mathrm{Var}}
	\!\left(\widehat{\mathcal C}_y^L\right)
}.
\end{equation*}
\end{document}